\documentclass[aps,prb,twocolumn,groupedaddress,showpacs]{revtex4}

\usepackage{amsmath}
\usepackage{amssymb}
\usepackage{epsfig}
\newcommand {\dr}{{\mathrm d}\mathbf{r}}

\newcommand {\dd}{{\mathrm d}}
\newcommand {\rr}{\mathbf{r}}

\newcommand {\etal}{\begin{itshape}et al\end{itshape}.}

\begin{document}

\title{A model colloidal fluid with competing interactions:
bulk and interfacial properties}

\author{A.J. Archer$^1$}
\email{Andrew.Archer@bristol.ac.uk}
\author{D. Pini$^2$}
\author{R. Evans$^1$}
\author{L. Reatto$^2$}
\affiliation{1. H.H. Wills Physics Laboratory,
University of Bristol, Bristol BS8 1TL, United Kingdom\\
2. Dipartimento di
Fisica, Universit\`a degli Studi di Milano, Via Celoria 16, 20133 Milano, 
Italy}

\date{\today}

\begin{abstract}
Using a simple mean--field density functional theory theory (DFT),
we investigate the structure and phase behaviour of a model colloidal
fluid composed of particles interacting via a pair potential which has a hard
core of diameter $\sigma$, is attractive Yukawa at intermediate separations and
repulsive Yukawa at large
separations. We analyse the form of the asymptotic decay of
the bulk fluid correlation functions, comparing results from our DFT with
those from the self consistent Ornstein-Zernike approximation (SCOZA). In both
theories we find rich crossover behaviour, whereby the ultimate decay of
correlation functions changes from monotonic to long-wavelength
damped oscillatory decay on crossing certain lines in the phase diagram, or
sometimes from oscillatory to oscillatory with a longer wavelength.
For some choices of potential parameters we find, within the DFT, a
$\lambda$-line at which the fluid becomes unstable with respect to periodic
density fluctuations. SCOZA fails to yield solutions for state points near such
a $\lambda$-line. The propensity to clustering of particles, which is reflected
by the presence of a long wavelength $\gg \sigma$, slowly decaying oscillatory
pair correlation function, and a structure factor that exhibits a very sharp
maximum at small but non zero wavenumbers, is enhanced in states near the
$\lambda$-line. We present
density profiles for the planar liquid-gas interface and for fluids adsorbed
at a planar hard wall. The presence of a nearby $\lambda$-transition gives rise
to pronounced long-wavelength oscillations in the one-body densities at both
types of interface.
\end{abstract}


\maketitle

\section{Introduction}
Recently colloidal systems have been synthesised in which the
effective pair potential between the colloidal particles
is attractive just outside the
core separation distance, but is repulsive for larger particle
separations.\cite{GhezziEarnshawJPCM1997, SearetalPRE1999, Sedgwick1, Sedgwick2,
Bartlett, Bartlett2}
The long range repulsion stems from the colloids being
charged and the short range attraction is generated by the depletion
mechanism,\cite{BarratHansen} arising
from the addition of non-adsorbing polymers to the
solution.\cite{FortinietalJPCM2005}
Such competing interactions can give rise to phase behaviour that can be
very different from that found in `simple' liquids. Theory and simulation
in both two and three dimensions for model systems with competing interactions
predicts that such interactions can give rise to
a state with undamped periodic density fluctuations, which
indicates a transition to cluster or striped
phases (microphase separation).\cite{AndelamnetalJCP1987, KendricketalEPA1988,
SearGelbartJCP1999, GroenewoldKegelJPCB2001,
GroenewoldKegelJPCM2004, ImperioReattoJPCM2004, SciortinoetalPRL2004,
MossaetalLangmuir2004, SciortinoetalJPCB2005, ImperioReattoJCP2006}
In the cluster phase the colloids are ordered in
such a way that there are assemblies containing tens or hundreds of particles
and large voids between the clusters, containing hardly any particles.
Similarly, in the striped phase in two dimensions the
particles are arranged in parallel stripes with low density regions
in-between the stripes.\cite{SearetalPRE1999, ImperioReattoJPCM2004, 
ImperioReattoJCP2006}
In three dimensions the stripes form
a gel--like network of elongated clusters.\cite{Bartlett, SciortinoetalPRL2004,
SciortinoetalJPCB2005}
This behaviour is most striking given that the pair
interactions between the particles are spherically symmetrical and suggests that
such systems could be important candidates
for developing self-assembling pattern forming materials. 
The liquid-vapour phase transition may be preempted by a transition to a
cluster or stripe phase, but
if the long-range repulsion is not sufficiently strong, no 
permanent clusters or stripes are present, and a liquid-vapour phase transition
is found. However, the liquid-gas coexistence curve is unusually flat in the
critical region.\cite{PinietalCPL2000}

Whilst the bulk phase behaviour has begun to be understood,
there are no detailed studies of how the long-range decay of the pair
correlation functions is affected by the competition between attraction and
repulsion and
very little is known about the properties of such fluids in inhomogeneous
situations where the average fluid density is non-uniform.
In the present paper we present a mean-field
density functional theory (DFT) for a simple model of such systems which
provides a means of analysing correlation functions and a
first step in elucidating the properties of inhomogeneous
fluids composed of particles with competing interactions.

The particular model fluid we consider in this paper is that described in
Refs.~\onlinecite{PinietalCPL2000, Pinietal2006, WuetalPRE2004,
LiuetalJCP2005, BroccioetalJCP2006}, in which particles
interact via a double-Yukawa pair potential of the form:
\begin{equation}
\beta v(r) = 
\begin{cases}
\infty \hspace{45mm} r \leq \sigma \\
-\epsilon \sigma \exp(-Z_1(r/\sigma-1))/r\\
 \hspace{3mm}+ A \sigma \exp(-Z_2(r/\sigma-1))/r \hspace{5mm} r > \sigma,
\end{cases}
\label{eq:pair_pot}
\end{equation}
where $\beta=1/k_BT$ is the inverse temperature (we shall set
$\beta=1$), $\sigma$ is the diameter of the particles and the amplitudes
$\epsilon>0$ and $A>0$. For a fixed value of $A$, the parameter $\epsilon$ plays
a role somewhat akin to an inverse temperature. In the context of
colloid-polymer mixtures one can envisage an athermal system at a fixed
temperature and varying the
chemical potential of a polymer reservoir thereby changing the strength of the
depletion attraction. Indeed, in the simple Asakura-Oosawa-Vrij model, the
depletion attraction between colloidal particles arising from the exclusion of
ideal polymers, yields a well depth proportional to $z_p$,
the fugacity of the ideal polymers.\cite{AO,Vrij,BraderetalMolecPhys2004}
Following the spirit of this approach we can view our parameter $\epsilon$ as a
measure of the polymer fugacity. In much of the work described
here we follow Pini \etal~\cite{PinietalCPL2000,Pinietal2006} by setting the
two (dimensionless) decay lengths $Z_1=1$ and $Z_2=0.5$. However, we also
present some results for other choices of $Z_1$ and $Z_2$.

A number of phase diagrams for this model system are displayed in
Refs.~\onlinecite{PinietalCPL2000} and \onlinecite{Pinietal2006},
for cases when $A$ is not too large,
i.e.\ cases when the repulsive contribution to the pair potential $v(r)$ is not
sufficiently large
that the liquid--gas phase separation is replaced by microphase
separation. For these cases one finds that there can be a substantial
supercritical region in the
phase diagram where the compressibility is unusually
large.\cite{PinietalCPL2000, Pinietal2006} In
Refs.~\onlinecite{PinietalCPL2000, Pinietal2006, WuetalPRE2004, LiuetalJCP2005,
BroccioetalJCP2006} a number of theories have been used to investigate
the bulk fluid correlation functions (radial distribution function $g(r)$
and structure factor $S(k)$). Here we make a systematic study of the
asymptotic decay $r \rightarrow \infty$ of $g(r)$ in various regions of the
phase diagram. The form of the decay of $g(r)$
is determined by the poles of $S(k)$ in the upper half of
the complex $k$--plane.\cite{paper6} We use a simple DFT,
equivalent to the random phase approximation (RPA) of the tail potential outside
the hardcore for the bulk fluid, to
elucidate the precise pole structure for particles interacting with
pair potentials given by Eq.~(\ref{eq:pair_pot}).
Due to the competing interactions the
fluid displays a propensity to cluster formation which is manifest for some
state points as a slowly decaying, long
wavelength oscillatory decay of the radial distribution function $g(r)$ -- the
wavelength being related to the size of the clusters. Of
course, at sufficiently low densities the decay of $g(r)$ is monotonic.
We locate lines in
the phase diagram at which the ultimate asymptotic decay of $g(r)$ crosses over
from monotonic to oscillatory. In addition we
find, for some values of the fluid pair
potential parameters, that there can be a crossover in the decay of $g(r)$ from
long wavelength oscillatory to oscillatory with wavelength $\sim \sigma$. We
find that the pole structure displayed by our simple (RPA) DFT, and hence the
crossover behaviour, is mimicked closely by
the more accurate self consistent Ornstein Zernike approximation 
(SCOZA),\cite{SCOZA,PinietalMolecP1998}
indicating that the simple DFT describes, at least
qualitatively, the main features of the fluid structure.

For sufficiently large values of $A$ we find that on increasing $\epsilon$, the
peak in the structure factor calculated from the DFT can diverge
at $k=k_c \neq 0$. We denote the line
in the phase diagram at which this divergence occurs the `$\lambda$-line'.
\cite{CiachetalJCP2003,Kirkwood,Archer6} On the $\lambda$-line the fluid becomes
unstable with respect to periodic density
fluctuations with wavenumber $k_c$, indicating that there
is a phase transition
to a modulated phase (either a crystalline or striped phase),
which may be pre-empted by a
transition to a glassy non--ergodic state.\cite{SciortinoetalPRL2004,
WuetalPRE2004,KroyetalPRL2004} In mean field treatments the
$\lambda$-line encloses a region of the phase diagram where one would
expect to find the liquid--gas critical point and it
intersects the binodal at densities on either side of the critical
point.\cite{KendricketalEPA1988}
We find such behaviour in the present DFT approach.

This paper is laid out as follows: In Sec.~\ref{sec:DFT} we introduce our simple
DFT approach and in Sec.~\ref{sec:SCOZA} we describe briefly the implementation
of the more accurate (bulk) SCOZA theory for determining correlation functions
and phase behaviour. In Sec.~\ref{sec:phase_diagrams}
we present results for the bulk phase
behaviour for a range of pair potential parameters.
Sec.~\ref{sec:ass} describes our approach for determining the ultimate
asymptotic decay of the radial distribution functions and presents a number of
representative results. In Sec.~\ref{sec:inhom_profiles} we apply our DFT to
calculate inhomogeneous fluid profiles at a planar hard wall and at the
liquid-gas planar free interface. We conclude,
in Sec.~\ref{sec:conc} with a summary and discussion of our results.

\section{A mean field density functional theory}
\label{sec:DFT}

The basis of DFT is that there
exists a functional, $\Omega_V[\rho]$, of the fluid one body density profile
$\rho(\rr)$ such that the fluid equilibrium profile $\rho(\rr)$ minimises this
functional
\begin{equation}
\frac{\delta \Omega_V[\rho']}{\delta \rho'(\rr)}
\Bigg \vert_{\rho'(\rr)=\rho(\rr)} = 0,
\label{eq:min_Omega}
\end{equation}
and the minimal value of this functional is the grand potential,
$\Omega$, of the fluid.\cite{Bob} The grand potential functional $\Omega_V$ can
be written as:
\begin{equation}
\Omega_V[\rho]={\cal F}[\rho]
-\int \dr \rho(\rr)[\mu-V_{ext}(\rr)],
\label{eq:Omega}
\end{equation}
where $\mu$ is the chemical potential, $V_{ext}(\rr)$ is a one--body external
potential and ${\cal F}[\rho]$, the intrinsic Helmholtz free energy functional,
is a unique functional of $\rho(\rr)$ for a given interaction potential. We can
write ${\cal F}[\rho]={\cal F}_{id}[\rho]+{\cal F}_{ex}[\rho]$ where
\begin{equation}
{\cal F}_{id}[\rho]=k_BT \int \dr \rho(\rr)[\ln(\rho(\rr)\Lambda^3)-1]
\label{eq:F_id}
\end{equation}
is the ideal--gas contribution, $\Lambda$ the thermal de Broglie
wavelength, and ${\cal F}_{ex}$ is the (unknown) excess contribution.
Taking two functional derivatives of ${\cal
F}_{ex}$, we obtain the (Ornstein-Zernike) pair direct correlation function for
the inhomogeneous fluid:\cite{Bob}
\begin{equation}
c^{(2)}(\rr,\rr')=
-\beta\frac{\delta^2 {\cal F}_{ex}[\rho]}{\delta \rho(\rr')\delta \rho(\rr)}.
\label{eq:c_2}
\end{equation}

\begin{figure}
\includegraphics[width=8cm]{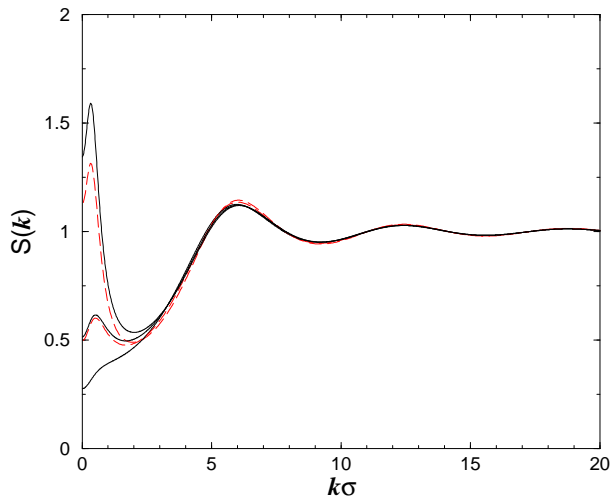}
\caption{Static structure factors $S(k)$ for a fluid with total
density $\rho_b\sigma^3=0.2457$, the critical point density determined from the
DFT (RPA), and with the parameters $A=0.082$, $Z_1=1$ and $Z_2=0.5$,
calculated for $\epsilon^{-1}=10$, 3, and 2 
(bottom to top). The solid lines are the results of the
DFT (or RPA), Eqs.~(\ref{eq:c2_fluid}) -- (\ref{eq:S_of_k}),
and the dashed lines from
SCOZA (there is no SCOZA result for $\epsilon^{-1}=10$). Note that as
$\epsilon^{-1}$ is decreased, a peak develops and grows at small wave vector
$k=k_c \neq 0$, whereas the other peaks in $S(k)$, determined primarily by
correlations in the reference hard sphere fluid, are almost unchanged.}
\label{fig:1}
\end{figure}

\begin{figure}
\includegraphics[width=8cm]{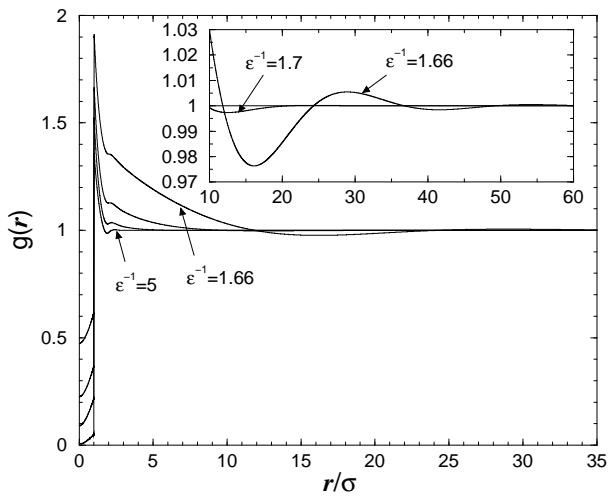}
\caption{Radial distribution function, $g(r)$, for a fluid with total
density $\rho_b\sigma^3=0.2$, with the parameters $A=0.082$, $Z_1=1$ and
$Z_2=0.5$, calculated for $\epsilon^{-1}=5$, 2, 1.7, and
1.66. Note that as $\epsilon^{-1}$ is decreased, $g(r)$
becomes longer ranged; the decay is
oscillatory, with a wavelength of about $25 \sigma$ for $\epsilon^{-1}=1.66$ (a
state close to the $\lambda$-line for this system -- see Fig.~\ref{fig:5})
-- see the magnification in the inset.
Within the present DFT (RPA) treatment, based on Fourier transforming
(\ref{eq:S_of_k}), the core condition $g(r)=0$ for $r<\sigma$ is
violated -- see text.}
\label{fig:2}
\end{figure}

In the present work we approximate the excess Helmholtz free energy
functional by
\begin{equation}
{\cal F}_{ex}[\rho] \,=\, {\cal F}_{ex}^{hs}[\rho]
\,+\, \frac{1}{2} \int \dr \int \dr'
\rho(\rr)\rho(\rr') v_p(|\rr-\rr'|),
\label{eq:freefunc_HS_ref}
\end{equation}
where ${\cal F}_{ex}^{hs}[\rho]$ is the reference hard-sphere Helmholtz excess
free energy functional and the contribution due to the remainder of the
pair potential $>\sigma$ is treated in a mean--field fashion.\cite{Bob} We
define the `perturbation' potential as
\begin{equation}
\beta v_p(r) = 
\begin{cases}
-\epsilon+A \hspace{5mm} r \leq \sigma \\
\beta v(r) \hspace{10mm} r > \sigma.
\end{cases}
\label{eq:v_p}
\end{equation}
Note that $-\epsilon+A$ is the value of $\beta v(r)$ at contact, i.e.~at
$r=\sigma^+$.
We employ the Rosenfeld fundamental measure theory
\cite{RosenfeldPRL1989, Rosenfeld:Levesque:WeisJCP1990, RosenfeldJCP1990}
for $F_{ex}^{hs}[\rho]$. This non-local functional, via (\ref{eq:c_2}),
generates $c_{PY}(r,\rho_b)$, the Percus-Yevick
(PY) pair direct correlation function for a uniform hard sphere fluid.
Thus, in bulk, Eqs.~(\ref{eq:c_2}) and
(\ref{eq:freefunc_HS_ref}) together generate the following simple
(random phase) approximation (RPA) for the pair direct correlation function:
\begin{equation}
c^{(2)}(r;\rho_b)\,=\, c_{PY}(r;\rho_b) \, - \, \beta v_p(r),
\label{eq:c2_fluid}
\end{equation}
where $\rho_b$ is the bulk fluid density. As in other RPA treatments of models
with hard cores the perturbation potential is not defined uniquely within the
hard core. We choose the form in Eq.~(\ref{eq:v_p}) for simplicity and because
the resulting phase diagrams are closer to those obtained from the SCOZA than
those from other possible choices that we considered.
With this choice the Fourier
transform of $c(r) \equiv c^{(2)}(r;\rho_b)$ can be carried out analytically and
yields
\begin{equation}
\hat{c}(k)\,=\, \hat{c}_{PY}(k) \, - \,\beta \hat{v}(k),
\label{eq:hat_c2_fluid}
\end{equation}
where $\hat{c}_{PY}(k)$, the Fourier transform of $c_{PY}(r;\rho_b)$,
is given by:\cite{Ashcroft:LecknerPhysRev1966}
\begin{eqnarray}
\hat{c}_{PY}(k)\,=&-&4 \pi \sigma^3 \bigg[ \left(\frac{\alpha+2 \beta+4
\gamma}{q^3}\,-\, \frac{24 \gamma}{q^5}\right) \sin q \notag \\
&+& \left( -\frac{\alpha+\beta+\gamma}{q^2} +
\frac{2 \beta + 12 \gamma}{q^4} - \frac{24 \gamma}{q^6} \right)
\cos q \notag \\
&+& \left( \frac{24 \gamma}{q^6} - \frac{2 \beta}{q^4} \right) \bigg]
\label{eq:hat_c2_fluid_PY}
\end{eqnarray}
with $q=k \sigma$, and the coefficients
\begin{eqnarray}
\alpha&=&\frac{(1+2\eta)^2}{(1-\eta)^4}, \notag \\
\beta&=&\frac{-6 \eta(1+\eta/2)^2}{(1-\eta)^4}, \notag \\
\gamma&=&\frac{\eta(1+2\eta)^2}{2(1-\eta)^4}
\end{eqnarray}
depend upon $\eta=\pi \rho_b \sigma^3/6$, the packing fraction, and
\begin{eqnarray}
\beta \hat{v}_p(k)=&-& \frac{4 \pi \sigma^3 \epsilon}{Z_1^2+q^2}
\left(\cos q+\frac{Z_1}{q}\sin q \right) \notag \\
&+& \frac{4 \pi \sigma^3 A}{Z_2^2+q^2}
\left(\cos q+\frac{Z_2}{q}\sin q \right) \notag \\
&-& \frac{4 \pi \sigma^3(\epsilon-A)}{q^3}
\left(\sin q-q\cos q \right).
\label{eq:hat_v_at}
\end{eqnarray}
Thus we have a relatively simple analytic expression for $\hat{c}(k)$ and
therefore also for the static structure factor obtained from the
Ornstein-Zernike relation:\cite{HM}
\begin{equation}
S(k)=\frac{1}{1-\rho_b \hat{c}(k)}.
\label{eq:S_of_k}
\end{equation}
In Fig.~\ref{fig:1}, we display $S(k)$ calculated at a bulk fluid
density $\rho_b \sigma^3=0.2457$, the critical point density as obtained from
DFT (see Sec.~\ref{sec:phase_diagrams}),
for a number of different values of the parameter $\epsilon$, for the
case when $A=0.082$, $Z_1=1$ and $Z_2=0.5$. We see that as
$\epsilon^{-1}$ is decreased, the structure factor develops a peak at a small,
but non-zero, wave-vector $k=k_c \ll 2\pi/\sigma$. This peak indicates the
propensity towards clustering in the
fluid.\cite{Bartlett, SearGelbartJCP1999, SearetalPRE1999,
ImperioReattoJPCM2004, SciortinoetalPRL2004, WuetalPRE2004, LiuetalJCP2005,
BroccioetalJCP2006} A similar RPA approximation was
used in Ref.~\onlinecite{ImperioReattoJPCM2004} to account for the
structure factor of a two dimensional fluid of particles with competing
interactions close to those of the present fluid. The propensity towards
clustering in the fluid can
also be seen in Fig.~\ref{fig:2} where we plot the radial distribution function,
$g(r)$, for the same fluid at $\rho_b \sigma^3=0.2$ and several values of
$\epsilon^{-1}$. We observe the development of long wavelength oscillations in
$g(r)$ as $\epsilon^{-1}$ is decreased. These results are obtained from the
inverse Fourier transform of $S(k)$.
Note that within this route, the core condition that $g(r)=0$ for
$r<\sigma$ is violated. This would not be the case if we were to use
the test particle
route to $g(r)$ -- i.e.~if we treat one of the particles as a fixed external
potential ($V_{ext}(\rr)=v(r)$ in Eq.~(\ref{eq:Omega})),
use the DFT to calculate the inhomogeneous
fluid density profile, $\rho(r)$, around this
fixed particle, and divide by the bulk density to obtain
$g(r)=\rho(r)/\rho_b$.

For sufficiently large values of $A$ we find that on
decreasing $\epsilon^{-1}$ at fixed $\rho_b$ the peak in the structure factor
that occurs at small $k=k_c \neq 0$
can diverge. The line in the phase diagram at which this occurs is the
$\lambda$-line.\cite{CiachetalJCP2003,Kirkwood,Archer6} This line
is shown in the inset of Fig.~\ref{fig:5} for the case $A=0.082$.
Inside the region bounded by the $\lambda$-line, the RPA
result for $S(k)$ is unphysical, since this becomes negative in a certain
interval of $k$, and goes from negative
to positive divergent values upon crossing the boundaries of the interval.

\section{The SCOZA for correlation functions and thermodynamics}
\label{sec:SCOZA}

The SCOZA is designed to deal with two-body potentials which, like that of
Eq.~(\ref{eq:pair_pot}), consist of a singular hard-sphere repulsive part, with
diameter $\sigma$, and
a longer-ranged tail. As is customary in integral-equation theories,
this approach is based upon the Ornstein-Zernike (OZ) equation linking the
radial distribution function $g(r)$ to the pair direct correlation
function $c(r)$. A closed theory is obtained by supplementing the OZ equation
with an approximate (closure)
relation involving $g(r)$ and $c(r)$. In its simplest
form, the SCOZA amounts to setting:\cite{SCOZA}
\begin{equation}
\left\{
\begin{array}{ll}
g(r)=0                     & r\leq \sigma \\
c(r)=K(\rho_b, \beta)v(r) \mbox{\hspace{0.3 cm}} & r>\sigma  \, .
\end{array}
\right.
\label{closure}
\end{equation}
This closure differs from the RPA considered in Sec.~\ref{sec:DFT} in two
respects: first, the core condition on the radial distribution function is
satisfied. Second, the amplitude $K$ of the direct correlation function 
outside the repulsive core has not been set to $K\equiv -\beta$.
Rather $K$ is regarded as an unknown state-dependent quantity,
to be determined in such a way that consistency between the compressibility
and the energy route to thermodynamics is enforced. 
The previous applications of SCOZA, including the study of
fluids with competing interactions,\cite{PinietalCPL2000,Pinietal2006}
were aimed at thermal systems with a temperature-independent potential,
such that the phase diagram can be
plotted as a function of density and temperature. 
For thermal fluids, the consistency condition  
amounts to requiring that the reduced compressibility $\chi_{\rm red}$ and
the excess internal energy per unit volume $u$ satisfy the condition:
\begin{equation}
\frac{\partial}{\partial \beta}\left(\frac{1}{\chi_{\rm red}}\right)=
\rho_b \frac{\partial^2 u}{\partial \rho_b^2} \, ,
\label{consist}
\end{equation}
where $\chi_{\rm red}$ is obtained from the compressibility
sum rule, i.e.\ $\chi_{\rm red}=S(0)$, while $u$ is
obtained from the energy equation
$u=2 \pi \rho_b^2 \int_{\sigma}^{\infty} \dd r\, r^2 v(r) g(r)$.
If the closure~(\ref{closure}) is implemented
for the correlations, the consistency
condition~(\ref{consist}) yields a closed partial differential equation
(PDE) for the function $K(\rho_b,\beta)$. 
It should be noted that in the present case 
$\beta v(r)$ is in effect independent of temperature
(see Eq.~(\ref{eq:pair_pot})), and
the phase diagrams will be plotted as 
a function of $\epsilon^{-1}$, the inverse strength of the attractive tail
potential.
For such an athermal system, Eq.~(\ref{consist}) does not hold anymore, 
as a consequence of the trivial dependence on the temperature of the 
Helmholtz free energy, $F\sim k_{B}T$. However, for any given $\epsilon$,
the value of $\beta F$ of the athermal system is the same as that given by 
a temperature-independent 
interaction $\widetilde{v}(r)=w\beta v(r)$
at a temperature $k_{B}T=w$, where $\beta v(r)$ is given 
in Eq.~(\ref{eq:pair_pot}), and $w$ is an arbitrary energy scale. 
For this system, Eq.~(\ref{consist}) is valid. Therefore, the phase
diagram and the correlations of the original potential (\ref{eq:pair_pot})
have been determined by integrating  
the SCOZA PDE for $\widetilde{v}(r)$ down 
to the reduced temperature $T^{*}=k_{B}T/w=1$ 
for each of the values of $\epsilon$ considered. 
Equivalently, one may reformulate Eq.~(\ref{consist}) for the athermal
system by replacing the inverse temperature $\beta$ with a coupling
constant which varies from zero to unity, whose purpose is to switch on 
the tail interaction.

In the present case, implementing
the SCOZA scheme is made simpler by taking advantage of the analytical
results~\cite{Hoyeetal} obtained for the mean spherical approximation (MSA)
when the tail potential has a two-Yukawa form as in
Eq.~(\ref{eq:pair_pot}). These enable one to obtain $\chi_{\rm red}$
as a function of $\rho_b$ and $u$. The function $\chi_{\rm red}(\rho_b,u)$
can then be used in Eq.~(\ref{consist}) by taking $u$ instead of
$K$ as the unknown quantity. The algebraic manipulations have been described 
in detail in Ref.~\onlinecite{PinietalJCP2001}.  
The resulting PDE for $u$ supplemented with suitable initial and boundary 
conditions~\cite{PinietalMolecP1998} is integrated numerically. 
Integration of $u$ with respect to $\beta$ then yields the Helmholtz free 
energy and hence all the other thermodynamic quantities. 

The SCOZA, like the RPA, yields a spinodal
curve, i.e., a locus in the temperature-density plane where 
the compressibility diverges. As soon as one enters the region bounded 
by the spinodal, $\chi_{\rm red}$ is no longer positive,
so that the theory ceases to be meaningful. Therefore, 
the region bounded by the spinodal must be excluded from the integration 
domain. If the temperature is below its critical value and one approaches
the critical density $\rho_{c}$
either from $\rho=0$ or from the high-density boundary 
$\rho_{0}$, it is found that $\chi_{\rm red}$ is no longer positive for
$\rho=\rho_{1s}$ or $\rho=\rho_{2s}$, where $\rho_{1s}$ and $\rho_{2s}$ 
are temperature-dependent densities with $\rho_{1s}<\rho_{c}<\rho_{2s}$.
The integration is then restricted to the intervals $(0,\rho_{1s}-\Delta\rho)$
and $(\rho_{2s}+\Delta\rho,\rho_{0})$, where $\Delta\rho$ is the density 
spacing. Within the accuracy of the numerical discretization, $\rho_{1s}$ and
$\rho_{2s}$ give the densites of the spinodal curve at the temperature 
considered. The liquid-gas coexistence curve is 
determined by equating the pressures and chemical potentials on the low- 
and high-density branches of the subcritical isotherms.

\section{Results for phase diagrams}
\label{sec:phase_diagrams}

\begin{figure}
\includegraphics[width=8cm]{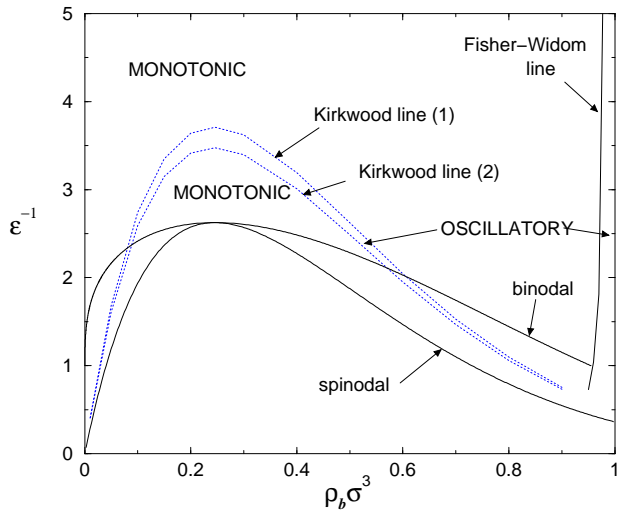}
\caption{Phase diagram for the case when $Z_1=1$, $Z_2=0.5$ and $A=0.0001$
obtained from the DFT (RPA). The solid lines denote the binodal and spinodal.
The two dotted
Kirkwood lines denote loci at which the asymptotic decay of
$h(r)$ crosses over from monotonic to oscillatory.
For the present choices of $Z_1$ and $Z_2$, the
Fisher-Widom line is at densities $\rho_b \sigma^3 \gtrsim 0.9$
which lie inside the liquid-solid coexistence region.\cite{Pinietal2006}}
\label{fig:3}
\end{figure}

\begin{figure}
\includegraphics[width=8cm]{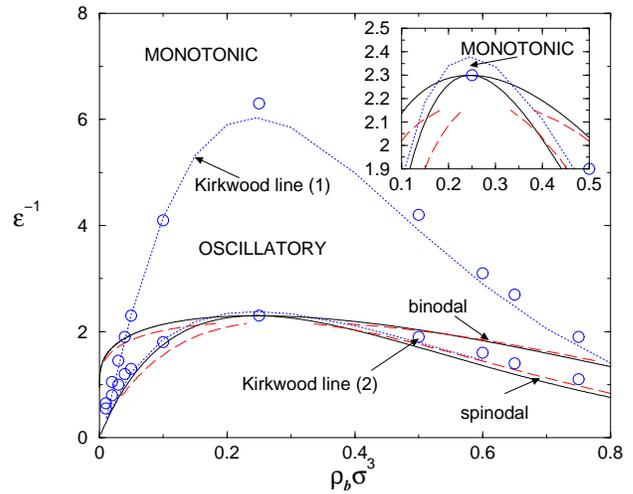}
\caption{As in Fig.~\ref{fig:3}, except here $A=0.02$ and we also display
SCOZA results: the dashed lines denote the SCOZA spinodal and binodal and the
open
circles correspond to points on the two SCOZA Kirkwood lines. Note that within
the DFT (RPA) the
lower Kirkwood (dotted) line is quite close to the
spinodal which means that there is only a very small supercritical region where
the asymptotic decay of $h(r)$ is monotonic. A similar scenario pertains with
SCOZA -- see magnification of the critical region in the inset.}
\label{fig:4}
\end{figure}

\begin{figure}
\includegraphics[width=8cm]{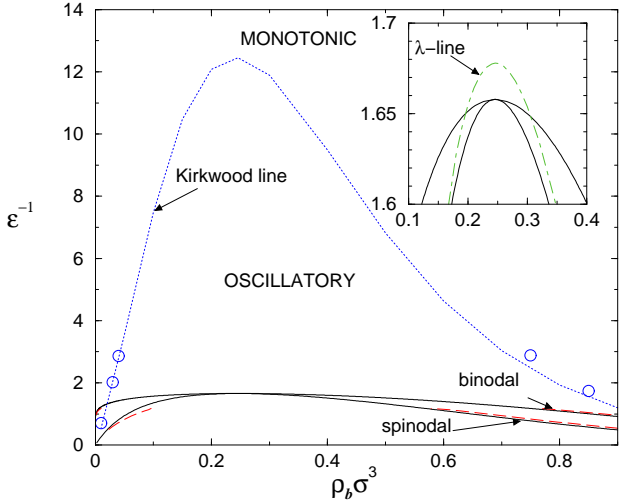}
\caption{As in Figs.~\ref{fig:3} and \ref{fig:4}, except here $A=0.082$.
The binodal and spinodal calculated from DFT (solid lines) lie close to the
corresponding SCOZA results (dashed lines). Note that solutions to SCOZA do not
exist for $1.18< \epsilon^{-1}<1.56$ and for $\epsilon^{-1}>4.04$ for this value
of $A$ -- see text. Within the DFT (RPA) there is a $\lambda$-line (dash-dotted
line in magnification in the inset) at which the
structure factor $S(k)$ diverges at $k=k_c \neq 0$, where $k_c \ll
2\pi/\sigma$.}
\label{fig:5}
\end{figure}

\begin{figure}
\includegraphics[width=8cm]{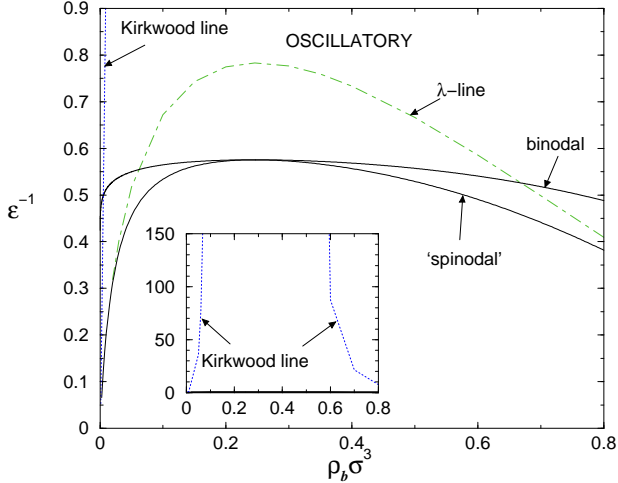}
\caption{As in Fig.~\ref{fig:3}, except here $A=0.5$. Now the region of the
phase diagram enclosed by the $\lambda$-line (dash-dotted)
is quite large. Furthermore, the
region enclosed by the Kirkwood (dotted) line is also very large --
see larger scale in the inset. Note that solutions to SCOZA exist only for
$\epsilon^{-1}<0.194$ -- see text.}
\label{fig:6}
\end{figure}

In this section we present results for the liquid-gas coexistence curve 
(binodal) and spinodal calculated from both theories. Within
the present DFT it is straightforward to determine
liquid-gas coexistence. By replacing $\rho(\rr)=\rho_b$ in the Helmholtz
free energy functional (\ref{eq:freefunc_HS_ref}),
we obtain an expression for the bulk Helmholtz free
energy per particle, $f$. The corresponding pressure can exhibit
a van der Waals loop and we calculate the coexisting gas and liquid
densities in the standard way. We obtain the spinodal from
the free energy as the locus of state points where
$\partial^2 f/\partial v^2=0$, where $v=1/\rho_b$
is the specific volume -- i.e.\ the locus where the isothermal
compressibility diverges. This is identical to the locus of points
where $S(k=0)$ diverges, where $S(k)$ is given by the Ornstein-Zernike relation
Eq.~(\ref{eq:S_of_k}), together with equations
(\ref{eq:hat_c2_fluid})--(\ref{eq:hat_v_at}). The self--consistency between the
free energy calculated directly from the functional and from the compressibility
route is one of the appealing features of the simple theory.
In Figs.~\ref{fig:3}--\ref{fig:6} we display the resulting DFT phase diagrams
for $Z_1=1$ and $Z_2=0.5$ and a
number of choices of the parameter $A$. In Figs.~\ref{fig:4} and \ref{fig:5} we
also display the corresponding
SCOZA results for the spinodal and binodal.
We find quite good agreement between the two theories, although the DFT being a
mean field theory, is unable to describe correctly the shape of the coexistence
curve in the region of the critical point. 

Within the present DFT, having
calculated the spinodal and binodal for a particular value of $A$, one can
obtain results for all other values of $A$ by a simple re--scaling of the
vertical ($\epsilon^{-1}$) axis. It is straightforward to show that,
within the DFT, the critical density is
independent of $A$. For the case $A=0$ (not displayed), the present theory
yields a critical density within 5\% of the
value obtained from the more accurate SCOZA.\cite{PinietalCPL2000}
However, as $A$ is increased the discrepancy
increases. The general trend that the value of $\epsilon^{-1}$ at the critical
point is decreased as $A$ is increased
is found in both theories.\cite{PinietalCPL2000}

In Figs.~\ref{fig:5} and \ref{fig:6}, which refer to $A=0.082$ and 0.5
respectively, the dash dotted line denotes the $\lambda$-line as obtained from
the DFT (RPA). 
The $\lambda$-line takes the shape of a loop which crosses both branches of
the binodal and meets the spinodal at
values of $\epsilon^{-1}$ below where one would expect the critical point.

When considering the SCOZA results, one should keep in mind that, unlike the
RPA, there are regimes in which the theory cannot be solved. 
In particular, this is the case for values of $A$ and $\epsilon$ such that 
the competition between attraction and repulsion is very strong,
and the fluid is expected to form microphases.  
In the RPA, this regime is marked by the appearance of
the $\lambda$-line, where the structure factor $S(k)$ has a singularity for 
$k=k_c\neq 0$. However, such a singularity
is incompatible with the fulfilment of the core condition~(\ref{closure}) 
if the direct
correlation function in Fourier space $\hat{c}(k)$ is analytic on the real 
axis, which is indeed the case in SCOZA. As a consequence,
the SCOZA fails to have solutions before a divergence in $S(k)$ occurs. 
Because of the non-local character of the SCOZA PDE, this lack of convergence
involves the full range of density. Therefore, is is not possible
to locate the $\lambda$-line using SCOZA. Another constraint is related 
to the numerical integration procedure, which requires the SCOZA 
PDE to be stable with respect to small fluctuations of the solution. 
However, such a requirement 
is not fulfilled when $A$ is sufficiently large that
the repulsive tail contribution to the potential overwhelms the attractive 
one. In this case, the PDE behaves like a diffusion equation with a negative
diffusion coefficient, and cannot be integrated numerically. An estimate 
of this intrinsic stability limit is provided by the condition that 
the spatial integral of the tail potential
$4 \pi \int_{\sigma}^{\infty} \dd r r^2 v(r)$
should remain negative.
How the above constraints affect the existence of the solution in the present
representation, in which the phase diagram is studied as a function of 
the attraction amplitude $\epsilon$ at fixed repulsion amplitude $A$, can be
gleaned by considering the case $A=0.082$ reported in Fig.~\ref{fig:5}. 
For a temperature-independent two-Yukawa potential with the same inverse-range
parameters $Z_{1}=1$, $Z_{2}=0.5$ as those considered 
here,\cite{PinietalCPL2000} the liquid-gas
transition is found to disappear when the relative amplitude of the repulsion
$A/\epsilon$ is larger than $0.097$. For $A=0.082$ this requires 
$\epsilon^{-1} > 1.18$. When this condition on $\epsilon$ is met, 
the theory fails to converge below
a certain threshold temperature $T_{th}$ before liquid-vapour separation takes
place. Whether this happens or not in the phase 
diagram of Fig.~\ref{fig:5} depends on whether the reduced temperature 
$T^{*}=1$ lies below or above the threshold temperature. 
For $A/\epsilon$ just above 0.097, Fig.~1 of Ref.~\onlinecite{PinietalCPL2000}
shows that $T_{th}^{*}\simeq 1.7\, \epsilon$ which, for 
$\epsilon^{-1}\simeq 1.18$, is indeed above unity. Therefore, 
the SCOZA solution
disappears as soon as $\epsilon^{-1} > 1.18$. However, as $\epsilon^{-1}$  
increases, the reduced threshold temperature $T_{th}^{*}$ decreases, until
for $\epsilon^{-1} > 1.56$ one has $T_{th}^{*}<1$, so that  
convergence for $T^{*}=1$ is again achieved. 
In the RPA picture, this corresponds to being above the maximum
of the $\lambda$-line. For larger values 
of $\epsilon^{-1}$, the SCOZA can be solved. However, the solution
disappears again at high values of $\epsilon^{-1}$, when 
the repulsive part
of the interaction dominates. For $Z_{1}=1$, $Z_{2}=0.5$, $A=0.082$, the SCOZA 
PDE fails to converge for $\epsilon^{-1} > 4.04$, in fair agreement with the
value
$\epsilon^{-1}=4.06$ above which the spatial integral of the tail potential 
for $r>\sigma$ becomes positive. In summary, for $A=0.082$ the SCOZA solution
does not exist for $1.18 < \epsilon^{-1} < 1.56$ and for $\epsilon^{-1}> 4.04$. 
Turning now
to the case $A=0.5$ shown in Fig.~\ref{fig:6}, the SCOZA solution disappears
for $\epsilon^{-1}>0.097/A=0.194$. However, for this choice of $A$ the solution
does not reappear at larger values of $\epsilon^{-1}$, because 
the region where the competition is strong and 
a $\lambda$-line is present according to RPA overlaps with the region 
$\epsilon^{-1}>0.67$ 
where the integral of the tail potential becomes positive and the SCOZA PDE
is unstable. Therefore, for $A=0.5$ only the interval $\epsilon^{-1}<0.194$ is 
accessible to the SCOZA. We did  
calculate portions of the binodal and spinodal in this interval,
finding good agreement with the DFT results. However, for these comparatively
low values of $\epsilon^{-1}$ the relevant densities are either  
very low or very high, and are not displayed in Fig.~\ref{fig:6}.  

\section{Asymptotic decay of correlation functions}
\label{sec:ass}

In this section we determine
the asymptotic decay, $r \rightarrow \infty$, of the
total pairwise correlation function $h(r)=g(r)-1$ in our model.
The basic procedure follows that in Ref.~\onlinecite{paper6}. In Fourier space 
the Ornstein-Zernike (OZ) relation is given by Eq.~(\ref{eq:S_of_k}), or
equivalently
\begin{equation}
\hat{h}(k)=\frac{\hat{c}(k)}{1-\rho_b \hat{c}(k)},
\label{eq:OZ}
\end{equation}
where $\hat{h}(k)$ is the three-dimensional Fourier transform (FT) of 
$h(r)$. Inverting the FT, and noting that $\hat{h}(k)$ is even, we can write:
\begin{eqnarray}
rh(r) &=& \frac{1}{4 \pi^2 i} \int_{-\infty}^{\infty} dk \, k \, e^{ikr}
\hat{h}(k)\notag \\
&=& \frac{1}{4 \pi^2 i} \int_{-\infty}^{\infty} dk \, k \, e^{ikr}
\frac{\hat{c}(k)}{1-\rho_b \hat{c}(k)},
\end{eqnarray}
which can be evaluated by contour integration in the complex plane.\cite{paper6}
We expect the singularities of ${\hat{h}}(k)$ to be simple poles.
Choosing an infinite 
radius semi-circle contour in the upper half of the complex plane, we obtain
\begin{equation}
r h(r) \, = \, \frac{1}{2 \pi} \sum_n R_n e^{i k_n r}
\label{eq:polesum}
\end{equation}
where $R_n$ is the residue of $k {\hat{h}}(k)$ for the $n$th pole at $k=k_n$. 
The $k_n$ are solutions of
\begin{equation}
1-\rho_b \hat{c}(k_n)=0.
\label{eq:pole_condition}
\end{equation}
In general, there are an
infinite number of poles. Poles lying off the imaginary axis occur in conjugate
pairs $k_n=\pm \alpha_1 + i\alpha_0$ and such a pair contributes a damped
oscillatory term of the form $\exp(-\alpha_0 r) \cos(\alpha_1 r-\theta)$
to the sum in Eq.~(\ref{eq:polesum}).
Poles that lie on the imaginary axis, $k_n=i
\tilde{\alpha}_0$, contribute a pure
exponential term of the form $\exp(-\tilde{\alpha}_0r)$ to the sum in
Eq.~(\ref{eq:polesum}).
The longest range decay of $h(r)$ is determined by the pole(s)
with the smallest imaginary part. If $\alpha_0 < \tilde{\alpha}_0$ the
longest range decay is damped oscillatory, but if $\alpha_0 > \tilde{\alpha}_0$,
then the asymptotic $r \rightarrow \infty$ decay of $h(r)$ is monotonic.

Given the analytic expression we have for $\hat{c}(k)$ within the DFT (RPA)
treatment (Eqs.~(\ref{eq:hat_c2_fluid})--(\ref{eq:hat_v_at})), it is fairly
straightforward to calculate the full set of solutions $\{k_n\}$ to
Eq.~(\ref{eq:pole_condition}) for the present model fluid.
Thus we are able to determine the pole(s) with
the smallest $\alpha_0$ and determine the ultimate decay
of $h(r)$.

Before presenting our numerical
results, we first discuss a simplified model, which
should provide insight into the origin of the leading order poles.
In Eq.~(\ref{eq:v_p}), $v_p(r)$ takes a constant value for $r<\sigma$.
If we replace this constant value by the form that $v(r)$ takes for
$r>\sigma$ (i.e.~keeping a double Yukawa form for {\em all} values of $r$) then
the denominator function $D(k) \equiv 1-\rho_b\hat{c}(k)$ takes the simple form:
\begin{equation}
D(k)=1-\rho_b \hat{c}_{PY}(k)
-\frac{4 \pi \sigma^3 \epsilon \rho_b}{Z_1^2+q^2}
+\frac{4 \pi \sigma^3 A \rho_b}{Z_2^2+q^2},
\label{eq:f}
\end{equation}
with $q=k \sigma$. We find that
this simplified model displays the same pole structure as the full version with
$v_p(r)$ given by Eq.~(\ref{eq:v_p}). When $A=0$ this simplifies even further:
\begin{equation}
D_0(k)=1-\rho_b \hat{c}_{PY}(k)-
\frac{4 \pi \sigma^3 \epsilon \rho_b}{Z_1^2+q^2}.
\label{eq:f_A=0}
\end{equation}
We first seek purely imaginary poles and substitute $k=i \tilde{\alpha}_0$
into Eq.~(\ref{eq:f_A=0}). Provided
$\tilde{\alpha}_0 \sigma \lesssim 10$, then
$\hat{c}_{PY}(i \tilde{\alpha}_0) \thickapprox \hat{c}_{PY}(0)$, so that:
\begin{equation}
D_0(i\tilde{\alpha})\thickapprox B- \frac{C}{Z_1^2-\sigma^2 \tilde{\alpha}_0^2},
\label{eq:f_A=0_a}
\end{equation}
where $B\equiv[1-\rho_b\hat{c}_{PY}(0)]>0$ and $C \equiv4 \pi \sigma^3
\epsilon \rho_b > 0$ are constants dependent on the state point of the fluid.
There is only one solution to the equation $D_0(i \tilde{\alpha}_0)=0$, where
$D_0$ is given by (\ref{eq:f_A=0_a}). When $B=C/Z_1^2$ then we obtain the
solution $\tilde{\alpha}_0=0$; this is just the spinodal. Furthermore,
the solution is bounded above, i.e.\ $\tilde{\alpha}_0 \sigma < Z_1$.

We find that there are an infinite
number of complex poles. These are essentially just the poles of the hard sphere
reference fluid. Of these, the pole with the
smallest imaginary part has a real part $\alpha_1 \thickapprox 2 \pi/ \sigma$.
The imaginary part of this pole, $\alpha_0$, takes
relatively large values at low densities and decreases as
the density is increased. Thus
we find, for the fluid with $A=0$, that at low densities the
purely imaginary pole dominates the asymptotic decay of $h(r)$.
However, as the density is
increased, one finds that at some point
$\tilde{\alpha}_0=\alpha_0$, and on further increasing the density one finds
that the asymptotic decay of $h(r)$ crosses over from monotonic
to damped oscillatory
decay, with a wavelength $\sim \sigma$. The line in the phase
diagram defined by the locus $\tilde{\alpha}_0=\alpha_0$ is termed the
Fisher--Widom (FW) line.\cite{paper6,FWline} For the case when $Z_1=1$ and
$Z_2=0.5$, this line is at high densities.
In Fig.~\ref{fig:3} we
display the FW line for the case $A=0.0001$ where this line is at densities
$\rho \sigma^3>0.9$ and lies within the liquid-solid coexistence
region.\cite{Pinietal2006}

We now consider the case $A \neq 0$. For purely imaginary poles,
$k=i \tilde{\alpha}_0$, Eq.~(\ref{eq:f}) becomes
(cf.~Eq.~(\ref{eq:f_A=0_a})):
\begin{equation}
D(i\tilde{\alpha})\thickapprox B- \frac{C}{Z_1^2-\sigma^2 \tilde{\alpha}_0^2}
+ \frac{E}{Z_2^2-\sigma^2 \tilde{\alpha}_0^2},
\label{eq:f_Aneq0}
\end{equation}
where $E \equiv 4 \pi \sigma^3 A \rho_b > 0$ is independent of
$\tilde{\alpha}_0$. When
$B=C/Z_1^2-E/Z_2^2$ then we obtain the solution $\tilde{\alpha}_0=0$ to the
equation $D(i\tilde{\alpha}_0)=0$, which corresponds to the spinodal. When $A$
is sufficiently small that there is no $\lambda$-line in the phase diagram,
then we find the following scenario:
At the critical point (or more generally, on
the spinodal) there are two roots of $D(i\tilde{\alpha}_0)=0$,
the solution $\tilde{\alpha}_0=\tilde{\alpha}_0^a=0$ and another at a larger
value of $\tilde{\alpha}_0=\tilde{\alpha}_0^b<Z_2/\sigma$. On increasing
$\epsilon^{-1}$, moving away from the spinodal, $\alpha_0^a$ increases in
value, whilst $\alpha_0^b$ decreases in value.
As one moves further from the spinodal these two roots (poles) coalesce at a
minimum of $D(i\tilde{\alpha}_0)$
in the interval $0<\tilde{\alpha}_0<Z_2/\sigma$. On moving still further
from the spinodal, we find that there are
no solutions to the equation $D(i\tilde{\alpha}_0)=0$, i.e.~there are no purely
imaginary poles in this portion of the phase diagram. 
Near the spinodal, $D(i\tilde{\alpha}_0)<0$ for
$Z_2/\sigma<\tilde{\alpha}_0<Z_1/\sigma$. However, on moving
away from the spinodal, one finds that in this interval the
function increases and its maximum touches the axis $D(i\tilde{\alpha}_0)=0$. On
increasing $\epsilon^{-1}$ further the function crosses the
axis and there are two roots (purely imaginary poles) in the
interval $Z_2/\sigma<\tilde{\alpha}_0<Z_1/\sigma$. As
one continues to move away from the spinodal these two
poles separate and very far from the spinodal one finds that one pole,
$\alpha_0^{a'} \rightarrow Z_1^-/\sigma$ and the other pole $\alpha_0^{b'}
\rightarrow Z_2^+/\sigma$.

\begin{figure}
\includegraphics[width=8cm]{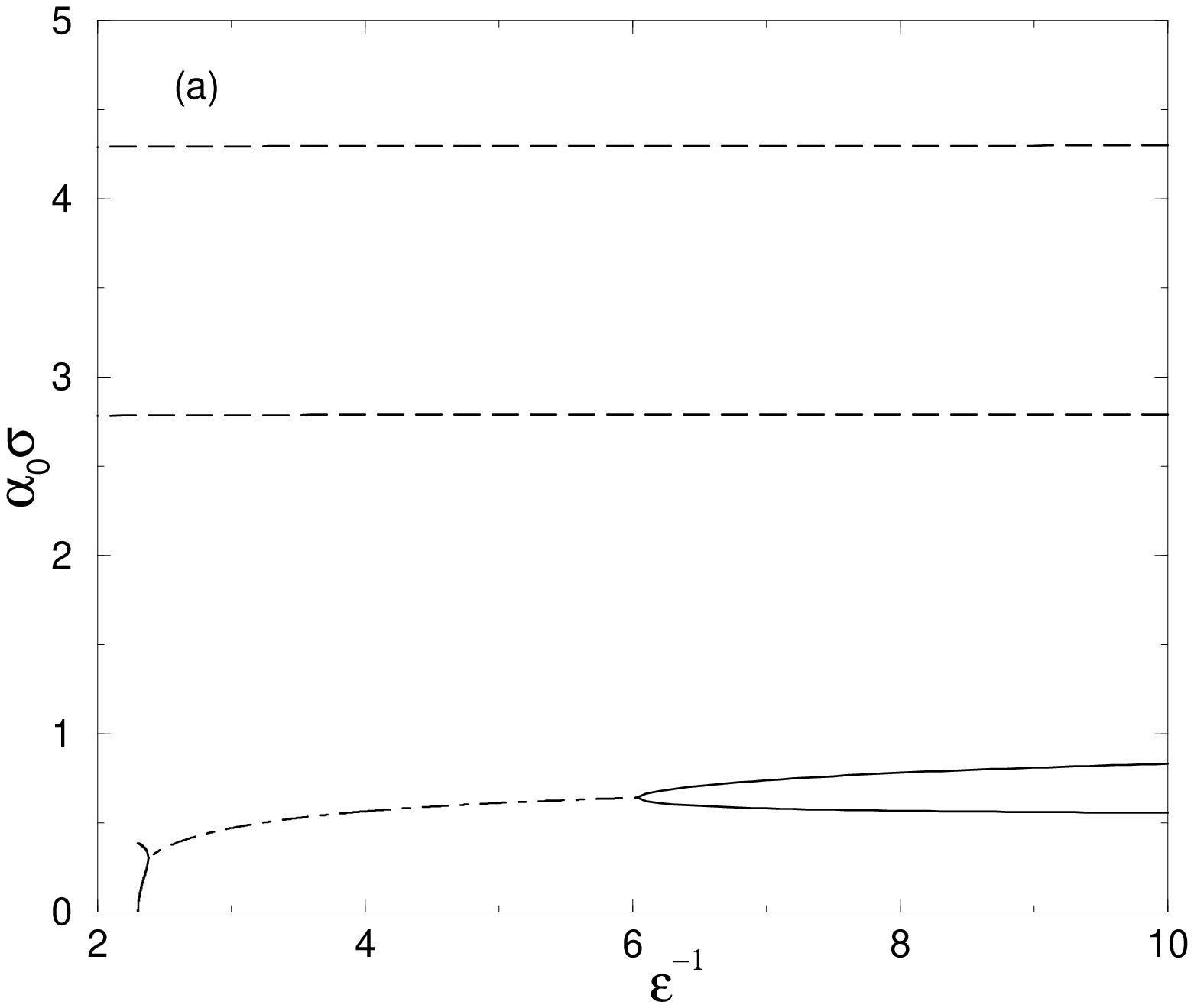}
\includegraphics[width=8cm]{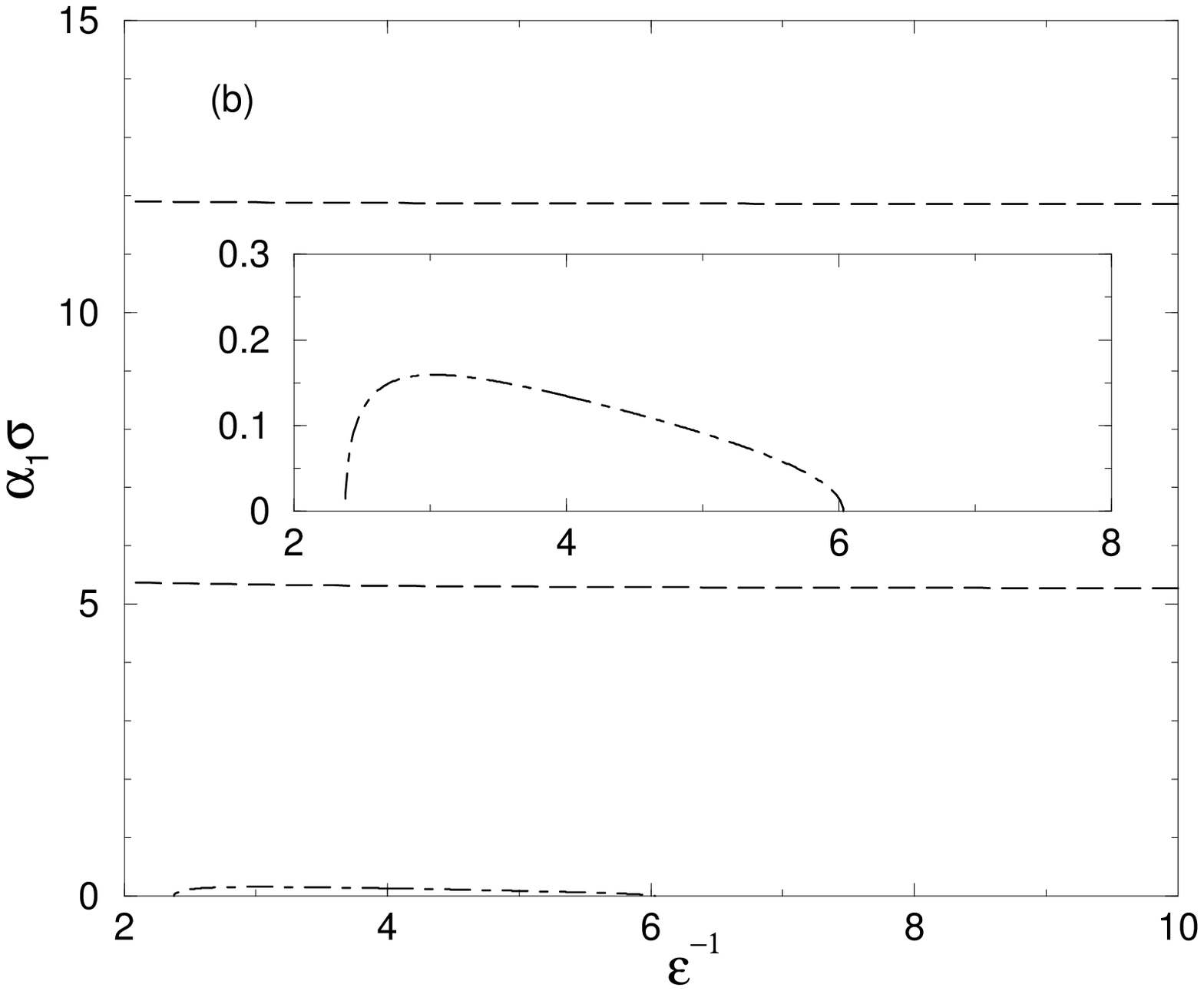}
\caption{(a) The imaginary part, $\alpha_0$, of the low lying poles for a fluid
with $A=0.02$, $Z_1=1$ and $Z_2=0.5$, calculated within the DFT (RPA)
along a line of constant
density $\rho \sigma^3=0.2457$, the critical density. The dashed lines
correspond to complex poles arising
from the reference hard--sphere correlations, the
dot dashed line to a complex pole with a small real part
$0<\alpha_1\ll2\pi/\sigma$ and the solid lines to two purely imaginary poles.
The point near $\epsilon^{-1}=2.38$ where the first pair of purely imaginary
poles coalesce corresponds to the first Kirkwood point and the point near
$\epsilon^{-1}=6.03$ where the second pair coalesce corresponds to the second
Kirkwood point for this density -- see Fig.~\ref{fig:4}. (b): The real part
$\alpha_1$ for the complex poles plotted
in (a). The inset is a magnification. Note that $\alpha_1 \sigma=0$ at the two
Kirkwood points and remains small in the oscillatory region.}
\label{fig:7}
\end{figure}

Analysing the roots of $D(k)=0$ more generally,
we find that when the two purely imaginary poles discussed above coalesce,
they form a conjugate
pair of complex poles at $k =\pm \alpha_1+ i \alpha_0$. At the point of
coalescence $\alpha_1=0$, i.e.\ the wavelength of the oscillations is infinite.
For all the relevant values of $\epsilon^{-1}$, $\alpha_1$ remains
small, such that $0<\alpha_1 \ll 2 \pi/\sigma$.
It is this conjugate pair of poles which generates the oscillatory
decay of $h(r)$ with a wavelength $\gg \sigma$,
indicating the tendency to cluster formation. The line
in the phase diagram at which these pairs of purely imaginary poles coalesce is
denoted the Kirkwood line.\cite{LeoDC:EvansMolP94,LeoDC:EvansMolP97,Hopkins}
Kirkwood\cite{Kirkwood} was the first to describe this mechanism for crossover
from monotonic to oscillatory decay in his study of (charge) correlations in
electrolytes. Moving away from the spinodal in the phase diagram, the conjugate
pair of complex poles with small $\alpha_1$ moves from the real axis
(increasing $\alpha_0$). Eventually $\alpha_1 \rightarrow 0$ and the poles
rejoin the imaginary axis at some point in the interval
$Z_2/\sigma<\tilde{\alpha}_0<Z_1/\sigma$, coalescing to form the second pair of
purely imaginary poles described above. There is therefore a second Kirkwood
line in the phase diagram. In Fig.~\ref{fig:7} we display the
poles calculated numerically from Eq.~(\ref{eq:pole_condition}) with
$\hat{c}(k)$ given by Eqs.~(\ref{eq:hat_c2_fluid})-(\ref{eq:hat_v_at})
along a path at the critical density, $\rho \sigma^3=0.2457$,
for a fluid with $A=0.02$, $Z_1=1$ and $Z_2=0.5$. These results exhibit all the
features of the simpler model described above.

Fig.~\ref{fig:7}a plots the imaginary part of the low lying poles. For small
values of $\epsilon^{-1}$ there is a pair of purely imaginary poles (solid line
in Fig.~\ref{fig:7}a) coalescing at the first Kirkwood point at
$\epsilon^{-1}=2.38$ and evolving as a conjugate complex pair (dash dotted line)
until the second Kirkwood point $\epsilon^{-1}=6.03$ when a second pair of pure
imaginary poles emerge. Fig.~\ref{fig:7}b shows that the real part of the lowest
lying complex pole $\alpha_1 \sigma$ remains $<0.16$ throughout the range where
the asymptotic decay of $h(r)$ is oscillatory. We see that the pole with
$\alpha_1 \thickapprox 2 \pi / \sigma$, arising from the reference hard sphere
correlations, has $\alpha_0 \sigma \thickapprox
2.8 >Z_1$ and therefore for all values of
$\epsilon^{-1}$, at this density, this pole does not determine the
asymptotic decay of $h(r)$. We also display the next higher order pole
arising from the
reference hard sphere correlations, which has $\alpha_1 \thickapprox 4
\pi/\sigma$ and $\alpha_0 \sigma \thickapprox 4.3$; the imaginary and real parts
of these poles are plotted as dashed
lines in Figs.~\ref{fig:7}a and \ref{fig:7}b.
Note that for the poles originating from the reference
hard sphere correlations the values of $\alpha_0$ and $\alpha_1$ change
very little as $\epsilon^{-1}$ is increased.

As the parameter $A$ is increased, the maximal separation between the two
Kirkwood lines increases (see Figs.~\ref{fig:3}--\ref{fig:5}); one Kirkwood
line moves towards the spinodal (decreasing $\epsilon^{-1}$)
and the other moves away from the spinodal.
Within the DFT the lower Kirkwood line meets the spinodal at a value of $A
\simeq 0.051$ (for $Z_1=1$ and $Z_2=0.5$). For greater values of $A$
a $\lambda$-line appears in the phase diagram,
enclosing the region where one expects the critical
point. The $\lambda$-line corresponds to the situation where the imaginary part
$\alpha_0 \rightarrow 0$ for the complex pole that has a
small real component
$0<\alpha_1 \ll 2 \pi/\sigma$ at some point above the spinodal. A divergence is
generated in the static structure factor at $k=k_c=\alpha_1$.
Decreasing $\epsilon^{-1}$ and following the $\lambda$-line, one finds that
the value of $k_c$ decreases continuously and eventually $k_c \rightarrow 0$,
i.e.\ the $\lambda$-line converts to the spinodal.
For the case $Z_1=1$ and $Z_2=0.5$ 
we could not obtain a solution for the SCOZA, near where one might expect
to find the $\lambda$-line, above a certain value of $A$.
On the basis of the discussion at the end of 
Sec.~\ref{sec:phase_diagrams}, 
the value of $A$ above which SCOZA fails to converge is determined 
by requiring that the threshold value 
of the relative amplitude\cite{PinietalCPL2000} 
$A/\epsilon=0.097$ is reached when the corresponding 
threshold temperature $T_{th}^{*}\simeq 1.7\epsilon$ is equal to unity. 
This gives $A \gtrsim 0.057$. This bound on $A$ is somewhat
larger than the value $A=0.051$ predicted by the present DFT.

We confirmed that the scenarios for the pole structure described in
Fig.~\ref{fig:7} are also present in the SCOZA
theory. The calculation of poles is straightforward because in SCOZA one
has analytic expressions for $\hat{c}(k)$.\cite{Hoyeetal}
In Figs.~\ref{fig:4} and
\ref{fig:5} we display the Kirkwood lines (open circles)
calculated from
the SCOZA alongside those from DFT (dotted lines).
We find good agreement between the results, demonstrating that our simple DFT
(RPA) describes correctly the behaviour of the poles in the model fluid.

\begin{figure}
\includegraphics[width=8cm]{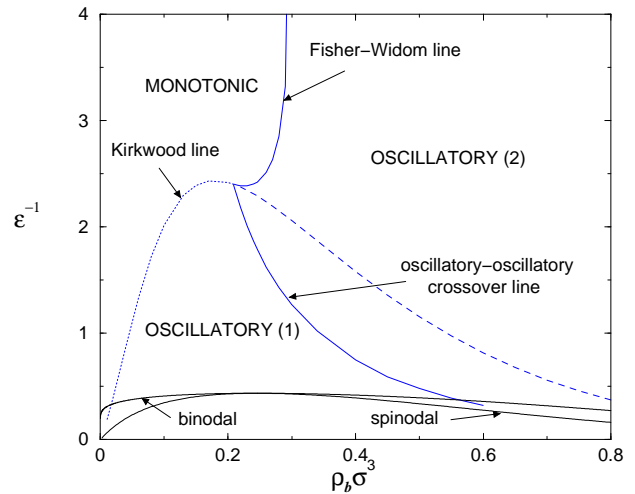}
\caption{Phase diagram for the case when $Z_1=6$, $Z_2=2$ and $A=0.3$. The
Kirkwood line (dotted) denotes the line at which the asymptotic decay of
$h(r)$ crosses over from monotonic to damped
oscillatory with wavelength $\gg \sigma$
and the FW line the crossover from monotonic to damped oscillatory with
wavelength $\sim \sigma$. Between the two regions of oscillatory
decay there is another (oscillatory to oscillatory) crossover line.
The dashed line is the continuation of the
Kirkwood line, although in this region of the phase diagram it denotes crossover
between higher order poles, i.e.\
those with larger values of $\alpha_0$ than the
leading order pole. There is also a second Kirkwood line located just above the
critical point so that in the immediate vicinity of the critical point $h(r)$
decays monotonically (OZ-like). However, on the scale of this figure, this
second Kirkwood line is indistinguishable from the spinodal.}
\label{fig:8}
\end{figure}

\begin{figure}
\includegraphics[width=8cm,height=5.3cm]{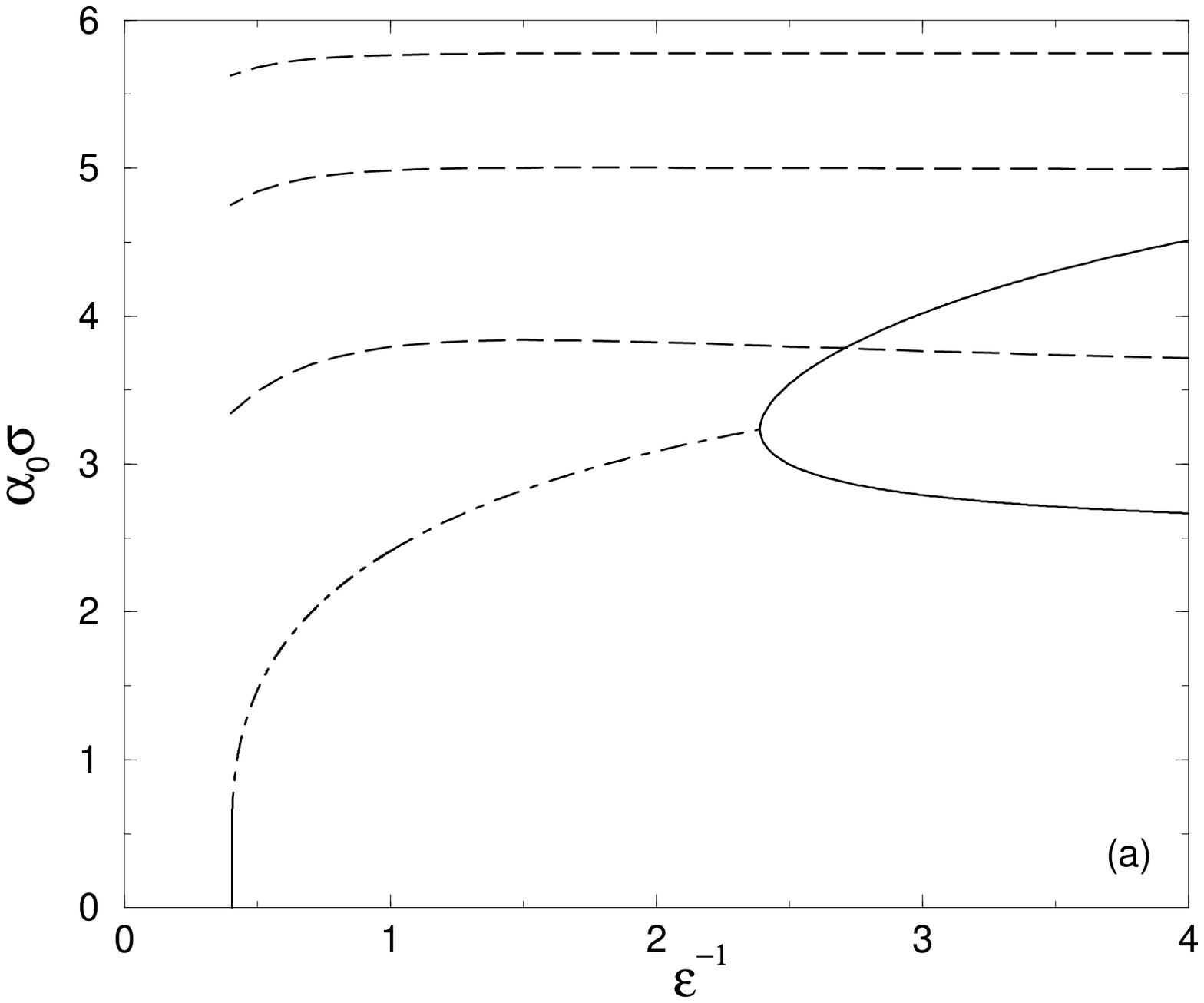}
\includegraphics[width=8cm,height=5.3cm]{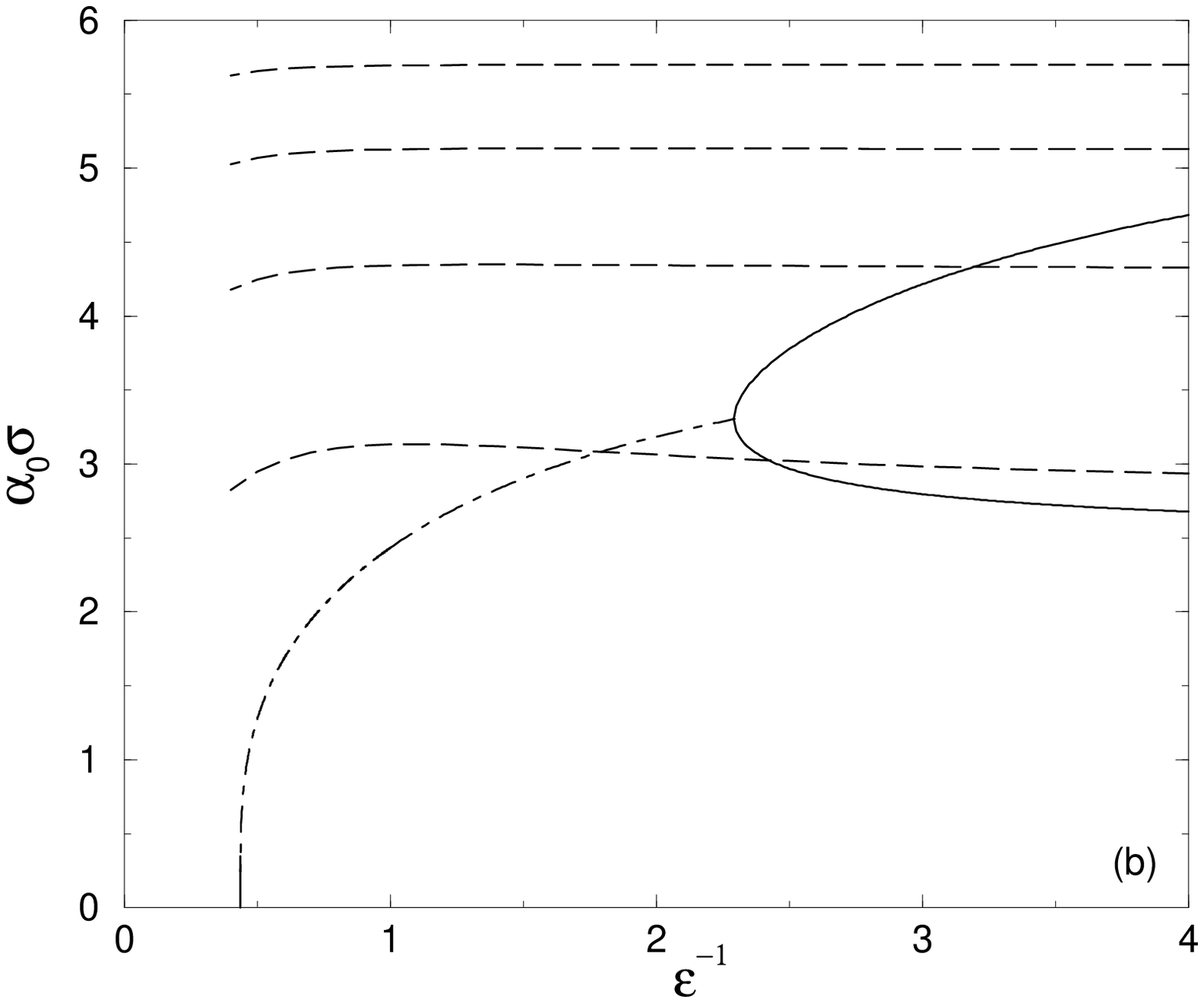}
\includegraphics[width=8cm,height=5.3cm]{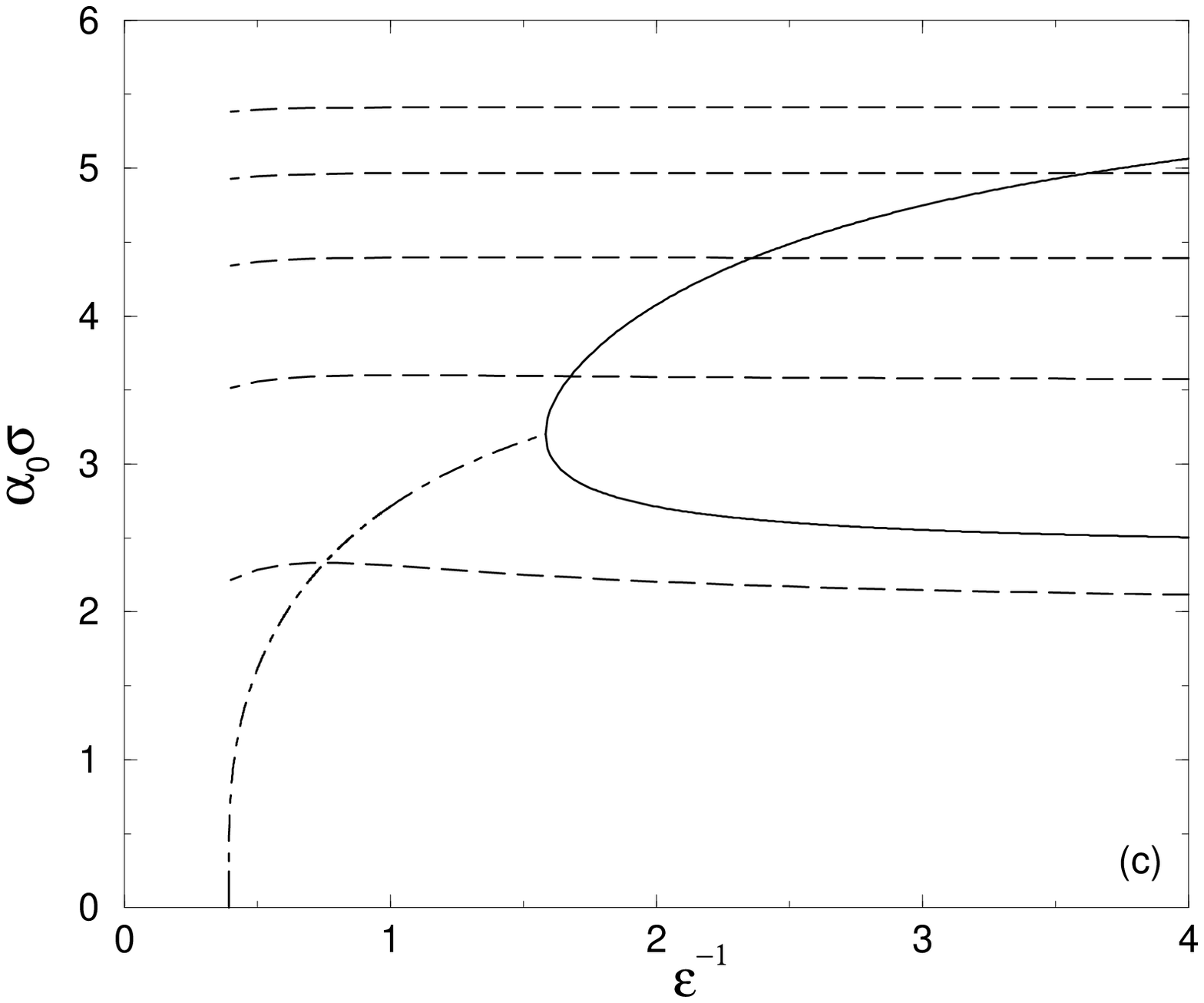}
\caption{The imaginary part, $\alpha_0$, of the lowest lying poles for a fluid
with $Z_1=6$, $Z_2=2$ and $A=0.3$, calculated within the DFT (RPA)
along lines of constant
density: (a) $\rho\sigma^3=0.15$, (b) the critical density
$\rho\sigma^3=0.2457$ and (c)
$\rho\sigma^3=0.4$. The dashed lines correspond to complex poles
arising from the reference hard--sphere correlations, the
dot dashed line to the complex pole with a small real part
$0<\alpha_1\ll2\pi/\sigma$, and the two solid lines to
two purely imaginary poles. The point where the two lowest lying pure imaginary
poles coalesce corresponds to a Kirkwood point -- see (a). In (b) this
coalescence occurs between higher order poles. The crossover at
$\epsilon^{-1}=1.79$ is between complex poles with different oscillatory
wavelengths -- see the line in Fig.~\ref{fig:8}. The subsequent crossover at
$\epsilon^{-1}=2.42$ is from oscillatory to monotonic decay. In (c) there is
oscillatory to oscillatory crossover at $\epsilon^{-1}=0.75$. The pole with
$\alpha_1 \sigma \sim 2 \pi$ remains the dominant one for larger
$\epsilon^{-1}$. Note that in (a)-(c) there is a
second Kirkwood point at $\epsilon^{-1}\thickapprox 0.4$, very near to the
critical value of $\epsilon^{-1}$, so that at smaller values of
$\epsilon^{-1}$ (close to the spinodal) there are two purely
imaginary poles, similar to the results displayed in Fig.~\ref{fig:7}a.}
\label{fig:9}
\end{figure}

We emphasise that there
are two distinct mechanisms for a crossover in the asymptotic decay of
$h(r)$ from monotonic to damped oscillatory: (i) the
coalescence of two purely imaginary poles to form a conjugate pair of complex
poles defines a point on the Kirkwood line and (ii) the simpler
Fisher--Widom (FW) mechanism, whereby on decreasing the
fluid density a purely imaginary
pole descends down the imaginary axis and acquires an imaginary part
$\tilde{\alpha}_0$ smaller than that of the dominant complex pole with imaginary
part $\alpha_0$, so that the ultimate decay becomes monotonic.
As mentioned earlier, when $Z_1=1$ and $Z_2=0.5$, the FW line is
at high densities, inside the region where we expect the solid phase to be the
equilibrium state -- see Fig.~\ref{fig:3}. However, for larger values of
$Z_1$ and $Z_2$ we
find that the FW line moves to lower densities, where it can intersect the
Kirkwood lines. In Fig.~\ref{fig:8} we display the phase diagram for a fluid
with $Z_1=6$, $Z_2=2$ and $A=0.3$. We find that when the Kirkwood and FW lines
intersect, another kind of structural crossover line appears in
the phase diagram: there is a crossover from damped oscillatory decay with
one wavelength to damped oscillatory decay with another
wavelength.\cite{Archer1,Grodon} In the present
case the crossover is from decay with a long wavelength of many particle
diameters (clustering in the fluid) to a decay with a wavelength $\sim \sigma$
as $\epsilon^{-1}$ is increased,
see Figs.~\ref{fig:8} and \ref{fig:9}. Note that for the choice of parameters
corresponding to the phase diagram in Fig.~\ref{fig:8}, the liquid-gas
transition may be weakly metastable. However, this does not
affect the decay crossover behaviour, which is in a region of the phase diagram
far from any phase transitions.\cite{Pinietal2006}

The pole analysis described above points to those regions of the phase
diagram where clustering
might occur, i.e.\ the region between the Kirkwood lines.
However, so far we have determined only the form of the
asymptotic decay of $h(r)$: whether it is monotonic or oscillatory as $r
\rightarrow \infty$. One must also calculate the amplitude of the
longest wavelength oscillatory contribution to $h(r)$ in order to assess
how significant clustering is in the fluid. This is straightforward within
the present DFT. The amplitude is determined by the residue of the pole (see
Eq.~(\ref{eq:polesum})), which is given by:
\begin{equation}
R_n=\frac{-k_n \hat{c}(k_n)}{\rho_b \hat{c}'(k_n)},
\label{eq:residue}
\end{equation}
where the prime denotes the derivative with respect to $k$.
For states just below the upper
Kirkwood line the amplitude is generally quite small.
For example, when $A=0.02$, for the fluid with $Z_1=1$ and $Z_2=0.5$ (see phase
diagram in Fig.~\ref{fig:4}), at the state point with $\rho_b \sigma^3=0.25$ and
$\epsilon^{-1}=5$, there is a complex pole with $\alpha_1\sigma = 0.0909$ and
$\alpha_0\sigma = 0.613$. This makes a damped oscillatory
contribution to $h(r)$ of the form ${\cal A} \exp(-\alpha_0 r) \cos(\alpha_1 r
-\theta)/r$ with amplitude ${\cal A}=0.0994$ and phase $\theta=0.992$, i.e.\ 
this term gives quite a small contribution to $h(r)$, despite being the term
that determines the ultimate
asymptotic decay. However, further below the Kirkwood line
the amplitude ${\cal A}$ can become larger and in the vicinity of a
$\lambda$-line, ${\cal A}$ can be very large so that the pole provides a large
contribution to $g(r)=1+h(r)$
-- see for example the results in Fig.~\ref{fig:2}.
Indeed, when the $\lambda$-line is reached, the amplitude ${\cal A}$ 
diverges
since the pole $k_{n}$ in Eq.~(\ref{eq:residue}) corresponds to a maximum
of $\hat{c}(k)$ on the real axis and $\hat{c}'(k_{n})$ vanishes. We will
return to this point in Sec.~\ref{sec:conc}.

\section{Inhomogeneous fluid: density profiles at interfaces}
\label{sec:inhom_profiles}

\begin{figure}
\includegraphics[width=8cm]{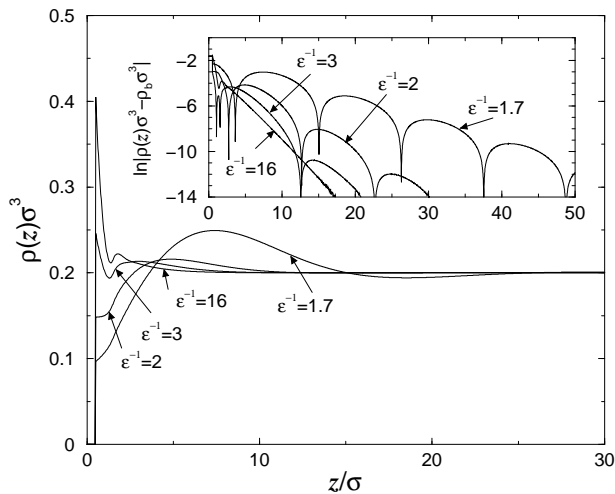}
\caption{Density profiles $\rho(z)$ at a planar hard wall for a fluid with
parameters $A=0.082$, $Z_1=1$ and $Z_2=0.5$ -- see phase diagram in
Fig.~\ref{fig:5}. The profiles are calculated for a fixed bulk density 
$\rho_b\sigma^3=0.2$, for $\epsilon^{-1}=1.7$, 2, 3 and 16. In the inset we plot
$\ln|\rho(z)\sigma^3-\rho_b\sigma^3|$. For $\epsilon^{-1}=16$ the asymptotic
decay of $\rho(z)$ is monotonic (exponential). The other three states lie on the
oscillatory side of the Kirkwood line. Note that the inverse
decay length increases as $\epsilon^{-1}$ is reduced and the $\lambda$-line is
approached. For a given value of $\epsilon^{-1}$ the wavelength of
the oscillations is the same as that of  $g(r)$ in Fig.~\ref{fig:2}.}
\label{fig:10}
\end{figure}

\begin{figure}
\includegraphics[width=8cm]{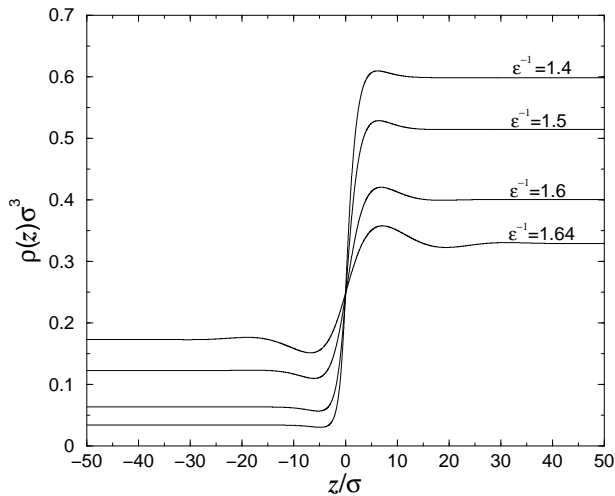}
\caption{Density profiles for the free gas--liquid interface calculated for
various values of $\epsilon^{-1}$, for a fluid with
parameters $A=0.082$, $Z_1=1$ and $Z_2=0.5$ -- see phase diagram in
Fig.~\ref{fig:5}. This fluid has a $\lambda$-line close to where one expects
the critical point. We find that as $\epsilon^{-1}$ is increased
the density difference between the coexisting phases decreases and the profiles
become more structured, i.e.\ the oscillations become more pronounced on both
sides of the interface. This is opposite to what occurs in simple fluids and
reflects the the proximity of the coexisting states to the $\lambda$-line.}
\label{fig:11}
\end{figure}

In this section we turn attention away from the uniform fluid to situations
where the one-body density is spatially varying. We consider two separate cases:
a) the fluid adsorbed at a hard wall and b) the liquid-gas interface. In both
cases the pole analysis for the decay of $h(r)$ is
directly relevant to the decay of
inhomogeneous fluid one--body density profiles; the same
pole(s) which determine the asymptotic decay of $h(r)$, determine the asymptotic
decay of the one-body density profiles.\cite{paper6} For
density profiles which vary only in one (Cartesian) direction, e.g.~when the
external potential $V_{ext}(\rr) \equiv V_{ext}(z)$, then provided $V_{ext}(z)$
is sufficiently short ranged the longest range 
decay of the profile into bulk at $z=\infty$ takes the form:
\begin{equation}
\rho(z)-\rho_b \sim \rho_b {\cal A}_w \exp(-\tilde{\alpha}_0 z), \hspace{5mm} z
\rightarrow \infty
\label{eq:rhodecay1}
\end{equation}
when the pole with the smallest value of $\alpha_0$ is purely imaginary and
\begin{equation}
\rho(z)-\rho_b \sim \rho_b \tilde{{\cal A}}_w \exp(-\alpha_0 z)
\cos(\alpha_1 z-\theta_w), \hspace{5mm} z \rightarrow \infty
\label{eq:rhodecay2}
\end{equation}
when the pole with the smallest $\alpha_0$ is complex. The poles $k_n=i
\tilde{\alpha}_0$ or $k_n=\pm \alpha_1+i \alpha_0$ correspond to the bulk
density $\rho_b$.\cite{paper6}

Using the DFT described in Sec.~\ref{sec:DFT}, we have determined the
one--body density profiles $\rho(z)$ for the two cases mentioned above.
Minimizing our approximation for $\Omega_V[\rho]$ yields
an Euler-Lagrange equation for the
equilibrium density profile $\rho(z)$ which can be solved
by means of a standard iterative scheme.
In Fig.~\ref{fig:10}, we display the one--body density profiles
at a planar hard wall with
\begin{equation}
V_{ext}(z) = 
\begin{cases}
\infty \hspace{5mm} z < \sigma/2 \\
0 \hspace{6mm} z > \sigma/2.
\end{cases}
\label{eq:V_ext}
\end{equation}
The fluid has
parameters $A=0.082$, $Z_1=1$ and $Z_2=0.5$ (see phase diagram in
Fig.~\ref{fig:5}). The density profiles are calculated for a fixed
bulk density $\rho_b\sigma^3=0.2$, for 
$\epsilon^{-1}=16$, 3, 2 and 1.7. For $\epsilon^{-1}=16$ the
state point is above the Kirkwood line, and the bulk
asymptotic decay of $h(r)$ is monotonic of the
form in Eq.~(\ref{eq:rhodecay1}).
This is confirmed in the inset of Fig.~\ref{fig:10} which
plots $\ln|\rho(z)\sigma^3-\rho_b\sigma^3|$ versus $z$;
when the decay of the density profile is of the
form in Eq.~(\ref{eq:rhodecay1}) the decay of the profile in this representation
is a straight line.
The other profiles are calculated for state points inside the Kirkwood line,
with increasing proximity to the $\lambda$-line. From the inset to
Fig.~\ref{fig:10} we see that the decay of these profiles is indeed
of the form in Eq.~(\ref{eq:rhodecay2}). The profile for
$\epsilon^{-1}=1.7$, a state point near to the $\lambda$-line,
has a small decay
length and oscillates with a long wavelength of about $23 \sigma$.
The density profiles for state points in the vicinity
of the $\lambda$-line are significantly different from those far away. These
profiles should be compared with the radial distribution functions $g(r)$ in
Fig.~\ref{fig:2} where the same trends with $\epsilon^{-1}$ are observed.

It is striking that as $\epsilon^{-1}$ is reduced the contact density
$\rho(\sigma/2)$ reduces dramatically and for $\epsilon^{-1}=1.7$ there is a
region where the
density near the wall is significantly depleted below bulk out to distances of
about $4 \sigma$. That the density at contact should become smaller follows from
the hard-wall sum rule: $k_BT \rho(\sigma/2)=p$, where $p$ is the pressure of
the bulk fluid far from the wall. Decreasing $\epsilon^{-1}$ corresponds to
turning on more of the attractive interaction thereby decreasing $p$.

In Fig.~\ref{fig:11} we display the density profiles
for the free planar gas--liquid interface of the same fluid. Profiles are
calculated for coexisting states with
$\epsilon^{-1}=1.4$, 1.5, 1.6 and 1.64. This fluid exhibits
a $\lambda$-line enclosing the region of the phase diagram
where one expects the critical point, see the inset to Fig.~\ref{fig:5}. States
outside the $\lambda$-line, i.e.\ for $\epsilon^{-1} \lesssim 1.65$,
correspond to conventional coexistence between (disordered) liquid and gas.
However, as the four coexisting states that we consider lie {\em inside}
the Kirkwood line, within our mean field DFT we
expect the asymptotic decay of the density
profiles to be damped oscillatory on {\em both} sides of the interface.
This is indeed what we find. Such behaviour was found previously in DFT studies
of fluid-fluid interfaces of the binary Gaussian core model.\cite{Archer1}
What is striking about the present results and what makes them different from
results for other models is that as $\epsilon^{-1}$ is increased and
the density difference between the coexisting phases
decreases, the profiles become more structured at the interface. Normally, the
interfacial density profile becomes less structured when
the difference in density between the coexisting phases decreases.
The difference is due to the fact that the present system exhibits a
$\lambda$-line and the coexisting densities become closer to this line as
$\epsilon^{-1}$ is increased -- see Fig.~\ref{fig:5}. The proximity of the
$\lambda$-line implies long-wavelength oscillatory density profiles decaying
slowly into bulk on either side of the interface. It should also imply that the
propensity to cluster formation in both phases is stronger. Clustering should
manifest itself in highly structured interfacial profiles.

\section{Discussion}
\label{sec:conc}

Using a simple DFT and the SCOZA integral equation theory we have investigated
the bulk structure and phase behaviour
of a model colloidal fluid with competing interactions. In particular, we
have examined in detail the asymptotic decay of the radial
distribution function $g(r)$ and find
a rich variety of decay scenarios. The presence of clustering in the fluid,
which occurs for sufficiently
large values of the repulsion amplitude $A$ in the pair potential
Eq.~(\ref{eq:pair_pot}),
is reflected in a long wavelength ($\gg \sigma$) damped oscillatory
decay of $g(r)$. For small values of $A$ 
the region of the phase diagram 
in which the decay of $g(r)$ is long wavelength oscillatory is bounded above and
below by two Kirkwood lines -- see Figs.~\ref{fig:3} and \ref{fig:4}.
On these lines the asymptotic decay of $g(r)$
crosses over from damped oscillatory to monotonic (exponential) decay.
As $A$ is increased we find within the mean-field DFT (RPA) that
the lower Kirkwood line is replaced by a $\lambda$-line
(see Figs.~\ref{fig:5} and \ref{fig:6}),
indicating a transition to a phase with undamped
periodic density fluctuations -- e.g.\ a cluster or a stripe phase.
Note that the appearance
of a region in the phase diagram with long wavelength oscillatory decay of
correlation functions is not
restricted to fluids with competing interactions. Such behaviour is to be
expected in any system with competing
interactions.\cite{SeulAndelmanScience1995,NussinovetalPRL1999} In
Ref.~\onlinecite{NussinovetalPRL1999} the authors investigate the form of the
correlation functions in frustrated $O(n)$ spin systems. 
They find that as the strength of the frustration parameter (the analogue of
the parameter $A$ in the present system) is
increased, a region in the phase diagram opens up in which
the decay of spin-spin correlations
is oscillatory, crossing over above and below to regions of
monotonic decay, in much the same way as in the
present fluid system.

In general, the results of our simple mean field DFT (RPA) are
in good qualitative and sometimes quantitative agreement with the
more sophisticated (and more accurate) SCOZA theory.
This reflects the fact that
our focus has been on state points where the parameters $A,\epsilon \lesssim 1$.
Recall that $(-\epsilon+A)k_BT$ is the value of the pair potential at contact,
$r=\sigma^+$, so that we have focused on cases where the portion of the
pair potential that we treat in mean field fashion has an amplitude $\lesssim
k_BT$. We expect the present mean field DFT to be less reliable when
$(\epsilon-A)>1$, or when the width of the attractive portion of the pair
potential becomes narrow, i.e.~when $Z_1$ becomes very large. In the
present study we have, in the main, avoided these regimes; an exception is the
system described in Fig.~\ref{fig:8} where $Z_1=6$.

In Sec.~\ref{sec:inhom_profiles} we presented results from DFT
for inhomogeneous one-body fluid density profiles. The oscillatory
density profiles in Figs.~\ref{fig:10} and \ref{fig:11}
correspond to state points between the Kirkwood and the $\lambda$-line in the
phase diagram of Fig.~\ref{fig:5}, where the system
is a disordered fluid, albeit with a propensity towards cluster formation.
For state points in the vicinity of the $\lambda$-line we find some very
striking density profiles with pronounced long wavelength ($\gg \sigma$)
oscillations. It is important to enquire whether such behaviour pertains beyond
the mean-field (MF) treatment we present here.

Determining the existence or non-existence of a $\lambda$-line is a problem that
arises for other types of fluids, e.g.\ microemulsions near the transition to a
lamellar phase and smectic liquid crystals. Recently much effort has focused on
a possible $\lambda$-transition for charge ordering in the restricted primitive
model (RPM) in which the two species of ions are modelled by equal sized hard
spheres carrying equal and opposite charges.\cite{CiachetalJCP2003} It is well
known that in the RPM the location of the $\lambda$-line in the phase diagram
depends sensitively on the choice of (MF) approximation -- see
Ref.~\onlinecite{CiachetalJCP2003} and references therein. For example, a simple
DFT yields a $\lambda$-line of continuous transitions between a uniform
disordered phase and a charge ordered phase, whereas the more sophisticated MSA
predicts no $\lambda$-transition. For the former case Ciach
\etal\cite{CiachetalJCP2003} studied correlation functions in some detail and
found that on approaching the $\lambda$-line, from the disordered side, both the
decay length {\em and} the amplitude of the charge-charge pair correlation
function diverge as $\tau^{-1/2}$, where $\tau$ measures the `distance' of the
state point from the $\lambda$-line. The derivation of Ciach \etal\ is based on
a particular form for the (charge-charge) inverse correlation function. However,
their results generalise straight forwardly\cite{Ciachcom} to any fluid system
where the dominant poles approach the real axis at $Re(k_n)=k_c\neq 0$ and the
other poles are well removed. As this situation pertains in the present DFT
(RPA) treatment of our model fluid we expect that close to the
$\lambda$-transition
\begin{equation}
rh(r) \sim {\cal A} \exp(-\alpha_0 r) \sin(k_c r), \hspace{3mm} r \rightarrow
\infty
\label{eq:28}
\end{equation}
with $\alpha_0 \sim \tau^{1/2}$ and ${\cal A}\sim \tau^{-1/2}$, where again
$\tau$ measures the distance from the $\lambda$-line. This result arises
also in a simplified treatment based on the small-$k$ expansion of the direct
correlation function: $1/S(k)=1-\rho \hat{c}(k)\sim a+bk^{2}+ck^{4}$.
For $a>0$, $c>0$, the condition that $1/S(k)$ be positive definite for every 
$k>0$ is violated for $b<0$, $b^{2}-4ac\geq 0$ and the appearance 
of the $\lambda$-line corresponds to the marginal case $b^{2}-4ac=0$. 
As the $\lambda$-line is approached from the disordered side, one has  
$b<0$, $b^{2}-4ac\rightarrow 0^{-}$, which via contour integration,
gives Eq.~(\ref{eq:28}) with ${\cal A}\sim 1/\sqrt{4ac-b^{2}}$.
Clearly the MF treatment implies the unphysical result that pair correlations
are unbounded on approaching the $\lambda$-line. What occurs beyond MF
theory? For the RPM a field theoretic treatment,\cite{CiachPatsahanPRE2006}
based on the approach of Brazovskii,\cite{Brazovskii} shows that incorporating
fluctuations leads to the disappearance of the $\lambda$-transition. Rather a
first order transition to a charge-ordered (crystalline) phase occurs at a
temperature below that of the original MF $\lambda$-transition. (Note that the
$\lambda$-transition is absent in simulation studies of the continuum RPM.)
Moreover the charge-charge correlation function changes smoothly near the
original MF $\lambda$-transition; the decay length and amplitude vary
continuously.\cite{PatsahanCiach}

One might infer that the very striking long-ranged decay of $g(r)$ observed in
Fig.~\ref{fig:2} for $\epsilon^{-1}=1.66$, at a state point rather close to the
$\lambda$-line in Fig.~\ref{fig:5}, is an artifact of the MF treatment since
including fluctuation effects would remove the $\lambda$-transition,
replacing this with a first-order transition to some ordered state (cluster or
stripe phase, perhaps). For state points somewhat further from the
$\lambda$-line, such as $\epsilon^{-1}=2.0$, the long wavelength oscillations
still develop in $g(r)$ but these are more highly damped. Nevertheless, the
corresponding structure factor $S(k)$ exhibits a pronounced peak at a non-zero
wavenumber $k_c$ -- see Fig.~\ref{fig:1}. The fact that SCOZA yields a very
similar structure factor gives us some confidence in the results of the mean
field DFT (RPA) for such state points.  
On the other hand, unlike the RPA, SCOZA does not yield 
a $\lambda$-line, rather it fails to converge in the regime in which 
a $\lambda$-line is expected
according to the RPA. As the convergence limit is approached, the cluster peak
in $S(k)$ may reach values much larger than unity, but nevertheless the peak
height does not grow arbitrarily, since a singularity in $S(k)$ at nonvanishing 
$k$ is incompatible with the core condition Eq.~(\ref{closure}), which is
fulfilled in SCOZA. This can be regarded as an indication that 
the $\lambda$-transition will be removed when one goes beyond 
the MF approximation. It is tempting to introduce a rough and ready
criterion, analogous to the Hansen-Verlet criterion for freezing, that
establishes when we expect the MF treatment to fail or to imply the onset of
ordering. Such a criterion might be when the peak height $S(k_c) \sim {\cal
O}(10^1)$.

Returning to the one-body profiles, we note that none of the bulk statepoints in
Fig.~\ref{fig:10} for adsorption at the planar hard wall are particularly close
to the $\lambda$-line and we might expect our DFT results to be at least
qualitatively correct. However, in Fig.~\ref{fig:11} for the density profiles at
the planar liquid-gas interface it is clear that the striking {\em increase} in
structuring as $\epsilon^{-1}$ is increased is a direct consequence of the close
proximity of the $\lambda$-line. It is unlikely that such behaviour could
survive beyond MF. A possible scenario when fluctuations are included is the
phase diagram of Fig.~\ref{fig:4} where there are two Kirkwood lines and the
critical point lies below the lower line (2). The MF DFT would yield
monotonically decaying profiles on both sides of the interface for coexisting
states just below the critical point but for $1.9 \lesssim \epsilon^{-1}
\lesssim 2.25$ one would expect damped oscillatory profiles on both the liquid
and the gas side. For states with $\epsilon^{-1} \lesssim 1.9$, where the upper
Kirkwood line (1) intersects the gas binodal, the asymptotic decay on the gas
side would be monotonic. Of course, the MF DFT omits effects of capillary wave
like fluctuations that act to erode the amplitude of the oscillations; this
mechanism is
discussed in Ref.~\onlinecite{BraderetalMolecPhys2004} for the case of model
colloid-polymer mixtures. We believe that much more work is required to
understand the nature of the fluid correlations in such systems, particularly in
regions of the phase diagram where the present MF theory predicts
a $\lambda$-line. Indeed, some of us are already engaged in such studies using
Monte-Carlo simulations and various integral equation theories.
Of course, one could also employ DFT to investigate possible spontaneous
ordering, i.e.\ whether there are solutions corresponding to non-uniform
(e.g.\ periodic; stripe or cluster) phases that have a lower free energy than
those corresponding to a uniform phase.

We conclude by speculating
that since colloidal systems with competing interactions display modulated
structures, these systems will have interesting optical properties and may have
applications in display technologies. 

\section*{Acknowledgements}

A.J.A. is grateful for the support of EPSRC under grant number GR/S28631/01.
D.P and L.R. acknowledge support from the Marie Curie program of the
European Union, contract number MRTN-CT2003-504712. We have benefited from
illuminating correspondence with Professor Alina Ciach on the subject of
$\lambda$-transitions and the failings of mean-field treatments, and we are
grateful to Dr. Nigel Wilding for several helpful discussions.


\end{document}